\documentclass[twocolumn,showpacs,preprintnumbers,amsmath,amssymb]{revtex4}
\usepackage[dvips]{graphicx}
\usepackage{graphicx,color}
\usepackage{latexsym,epsfig,bm,times,psfrag,subfigure}
\usepackage{color}
\usepackage{dcolumn}                 
\usepackage[bookmarksnumbered,bookmarksopen,colorlinks,citecolor=blue,linkcolor=blue]{hyperref} %
\usepackage{color, soul}
\setulcolor{red}
\setstcolor{red}
\sethlcolor{green}

\begin{document}

\title{Full jet evolution in quark-gluon plasma and nuclear modification of jet production and jet shape in Pb+Pb collisions at 2.76~ATeV at the LHC
}

\author{Ning-Bo Chang}
\affiliation{Institute of Particle Physics and Key Laboratory of Quark and Lepton Physics (MOE), Central China Normal University, Wuhan, 430079, China}

\author{Guang-You Qin}
\affiliation{Institute of Particle Physics and Key Laboratory of Quark and Lepton Physics (MOE), Central China Normal University, Wuhan, 430079, China}

\date{\today}

\begin{abstract}

We study the evolution of full jet shower in quark-gluon plasma via solving a set of coupled differential transport equations for the three-dimensional momentum distributions of quarks and gluons contained in the full jets.
In our jet evolution equations, we include all partonic splitting processes as well as the collisional energy loss and transverse momentum broadening for both the leading and radiated partons of the full jets.
Combining with a realistic (2+1)-dimensional viscous hydrodynamic simulation for the space-time profiles of the hot and dense nuclear medium produced in heavy-ion collisions, we apply our formalism to calculate the nuclear modification of single inclusive full jet spectra, the momentum imbalance of photon-jet and dijet pairs, and jet shape function (at partonic level) in Pb+Pb collisions at 2.76~ATeV.
The roles of various jet-medium interaction mechanisms on the full jet modification are studied. 
We find that the nuclear modification of jet shape is sensitive to the interplay of different interaction mechanisms as well as the energies of the full jets.

\end{abstract}

\maketitle

\section{Introduction}
\label{sec:intro}

Jet quenching is one of the main tools to study the properties of the quark-gluon plasma (QGP) created in the experiments of relativistic heavy-ion collisions, such as those at the Relativistic Heavy-ion Collider (RHIC) and the Large Hadron Collider (LHC) \cite{Wang:1991xy, Qin:2015srf, Blaizot:2015lma}.
In heavy-ion collisions, the hard partons produced at very early time pass through the surrounding hot and dense QGP formed in these energetic collisions before hadonizing into hadrons.
During the propagation, the partonic jet shower may experience elastic and inelastic scatterings with the constituents of the medium.
The interaction between the hard jets and the soft medium usually reduce the energy of the propagating jet partons, which leads to the suppression of final observed hadron and full jet yield at high transverse momentum, as compared to the binary-collision-scaled proton-proton collisions.
Such phenomenon is usually denoted as jet quenching.
In the last decades, there have been tremendous theoretical/phenomenological studies \cite{Gyulassy:1993hr,Baier:1994bd,Zakharov:1996fv,Gyulassy:2000fs,Wiedemann:2000za,Arnold:2001ba,Guo:2000nz,Bass:2008rv,Armesto:2011ht,Burke:2013yra} and experimental measurements \cite{Adcox:2001jp,Adler:2002xw,Aamodt:2010jd,Abelev:2012hxa} on the suppression of single inclusive hadron spectra at high transverse momentum regime.
In recent years, the experimental reconstruction of full jets with a given cone size defined as $R=\sqrt{(\Delta \phi)^2 + (\Delta \eta)^2}$ from large fluctuating background became available in relativistic heavy-ion experiments \cite{Aad:2010bu,Chatrchyan:2011sx,Chatrchyan:2012gt,Aad:2014bxa,Chatrchyan:2012gw, Chatrchyan:2013kwa, Aad:2014wha}, which has stimulated intensive theoretical interest in the study of full jets and their nuclear modification in high-energy nuclear collisions \cite{Vitev:2009rd,Qin:2010mn,CasalderreySolana:2010eh,Lokhtin:2011qq,Young:2011qx,He:2011pd,Renk:2012cx,Ma:2013pha,Senzel:2013dta,Chien:2015hda,Milhano:2015mng, Dai:2012am,Wang:2013cia,Qin:2012gp, MehtarTani:2011tz,CasalderreySolana:2012ef,Blaizot:2013hx,Fister:2014zxa, Apolinario:2012cg,Zapp:2012ak,Majumder:2013re}.

The study of full jets and their nuclear modification in heavy-ion collisions is quite different from single hadron production, and is also expected to provide more detailed information about jet-medium interaction.
For jet quenching study using single inclusive hadrons, the main focus is the interaction of the leading partons with the medium constituents. 
In this case, the radiative inelastic interaction with the medium constituents is usually regarded as the main mechanism responsible for parton energy loss and the suppression of light parton/hadron production in heavy-ion collisions \cite{Wicks:2005gt,Qin:2007rn,Schenke:2009ik}, while elastic collisions usually provides complementary contribution for light flavor jet quenching (We note that for heavy flavors, the collisional process become much more important due to finite mass effect, especially at low transverse momentum regime \cite{Cao:2013ita, Cao:2015hia}). 
Full jets contain the information of both the leading and subleading hadrons. Therefore, in order to achieve a full understanding of the nuclear modification of full jets, one has to take into account the effect of the medium on the full jet shower, i.e., the leading parton as well as subleading partons.
On one hand, since some of the radiated partons can be still in the jet cone, the energy loss for the leading parton is not the energy loss of the full jet.
On the other hand, the radiated shower partons in full jets may interact with the constituents of the medium, and contribute to the energy lose of the full jet \cite{Qin:2010mn}. 
Also, the interaction of full jets with the medium may change the distribution of the energy and momentum inside the jet cone, the study of which may provide stringent test on our modeling jet-medium interaction in relativistic heavy-ion collisions. 

In this work, we formulate a set of coupled differential transport equations to calculate the evolution of the three-dimensional momentum distributions of the shower partons contained in the full jet during the propagation through the hot and dense medium. 
Our evolution equations take into account all the partonic splitting channels occurring in the inelastic collision processes, as well as the collisinal energy loss and transverse momentum broadening experienced by the shower partons via elastic collisions with the medium constituents. 
By turning on or off the corresponding terms in the evolution equations, we can study the effects of different jet-medium interaction mechanisms on the propagation and mofication of full jet shower. 
We keep track of both the energies and the transverse momenta of all shower partons in the full jet during the evolution, thus the medium modification of full jet energy as well as jet internal structure due to jet-medium interaction can be studied straightforwardly.

The article is organized as follows.  
In Sec.~\ref{sec:framework}, we introduce our framework for calculating the evolution of the shower partons' three-dimensional momentum distributions in the full jets during the propagation through the hot and dense medium. 
In Sec.~\ref{sec:results}, we solve the coupled differential transport equations on the background QGP medium produced in Pb+Pb collisions at 2.76~ATeV, and present our results for various full jet observables, including the suppression of single inclusive jet spectra, the momentum imbalance of photon-jet and dijet pairs, and the modification of jet shape function. 
In Sec.~\ref{sec:analysis}, we investigate the effects of different jet-medium interaction mechanisms on the energy loss of full jets and on the nuclear modification of jet shape function. 
The summary is given in Sec.~\ref{sec:summary}.

\section{Framework for full jet shower evolution in medium}
\label{sec:framework}

In our approach, we foucs on the energy and transverse momentum distributions of the shower partons inside the full jet, $f_i(\omega_i, k_{i\perp}^2)=dN_i(\omega_i, k_{i\perp}^2)/d\omega_i dk_{i\perp}^2$, with $\omega_i$ the parton $i$'s energy and $k_{i\perp}$ its transverse momentum with respect to the jet axis. 
This three-dimensional momentum distribution contain much information about how the energy and momentum are distributed within the full jet.
In particular, we calculate the total energy of the full jet with a given cone size $R$ as follows: 
\begin{eqnarray}
E_{jet}(R) = \sum_i \int_R \omega_i f_i(\omega_i, k_{i\perp}^2) d\omega_i dk_{i\perp}^2,
\end{eqnarray} 
where the subscript $R$ means that the integral is taken with the constraint, ${k_{i\perp}}< {\omega_i}R$. 
To study the medium effect on the propagation and the modification of the full jet shower, we solve a set of coupled-differential transport equations for the three-dimensional distribution $f_j(\omega_j, k_{j\perp}^2, t)$. The transport equations have the following generic form:
 \begin{eqnarray}
\label{eq:dG/dt2}
\!\!&&\!\!\frac{d}{dt}f_j(\omega_j, k_{j\perp}^2, t) = \left(\hat{e}_j \frac{\partial}{\partial \omega_j}
  + \frac{1}{4} \hat{q}_j {\nabla_{k_\perp}^2}\right)f_j(\omega_j, k_{j\perp}^2, t)  \ \ \ 
\nonumber \\
\!\!&&\!\!+\sum_i\int d\omega_idk_{i\perp}^2 \frac{d\tilde{\Gamma}_{i\rightarrow j}(\omega_j, k_{j\perp}^2|\omega_i, k_{i\perp}^2)}{d\omega_j d^2k_{j\perp}dt} f_i(\omega_i, k_{i\perp}^2, t)\nonumber\\
\!\!&&\!\!-\sum_i \int d\omega_idk_{i\perp}^2 \frac{d\tilde{\Gamma}_{j\rightarrow i}(\omega_i, k_{i\perp}^2|\omega_j, k_{j\perp}^2)}{d\omega_i d^2k_{i\perp}dt} f_j(\omega_j, k_{j\perp}^2, t).
\end{eqnarray}
In the above equation, the first and second terms on the right hand side account for the contributions from the collisional energy loss and transverse momentum broadening experienced by the shower partons during the interaction with medium constituents via binary elastic collisions. 
The sizes of these two contributions are controlled by two transport coefficients, the longitudinal momentum loss rate $\hat{e}=d\langle E\rangle/dt$, and the exchange rate of transverse momentum squared $\hat{q}=d\langle \Delta p_\perp^2\rangle/dt$ \cite{Baier:1996kr, Majumder:2007hx, Qin:2012fua}.
The remaining terms are the contributions from various inelastic processes. 
The second line is the gain term, accounting for the radiation of parton $j$ with the energy $\omega_j$ and the transverse momentum $k_{j\perp}$ from the parton $i$ with the energy $\omega_i$ and the transverse momentum $k_{i\perp}$. 
The third line is the loss term, representing the contribution from the radiation of parton $i$ from parton $j$. 
We note that for the radiation process $i \to j$, it contributes not only to the gain term for the distribution $f_j(\omega_j, k_{j\perp}^2)$, but also to the loss term for the distribution $f_i(\omega_i, k_{i\perp}^2)$. 
Our current framework is an extension of the approach used by one of the authors in Ref. \cite{Qin:2010mn}, but contains several improvements. 
For the contribution from the inelastic processes, we include the gluon radiation as well as $g \to q \bar{q}$ channel. 
Also in Ref. \cite{Qin:2010mn} the radiative contribution is only considered for the leading parton in the jet, while in the current work we take into account the effect that the radiated partons can also experience inelastic collisions with the medium consituents after they are formed. 
Our transport equation is similar to that used in Ref. \cite{Jeon:2003gi, Qin:2007rn}, but we now apply to the shower partons contained in the full jets and keep track of their energies as well as their transverse momenta during the evolution through the hot and dense nuclear medium. 

\begin{figure}[htbp]
  \centering
     \includegraphics[width=0.48\textwidth]{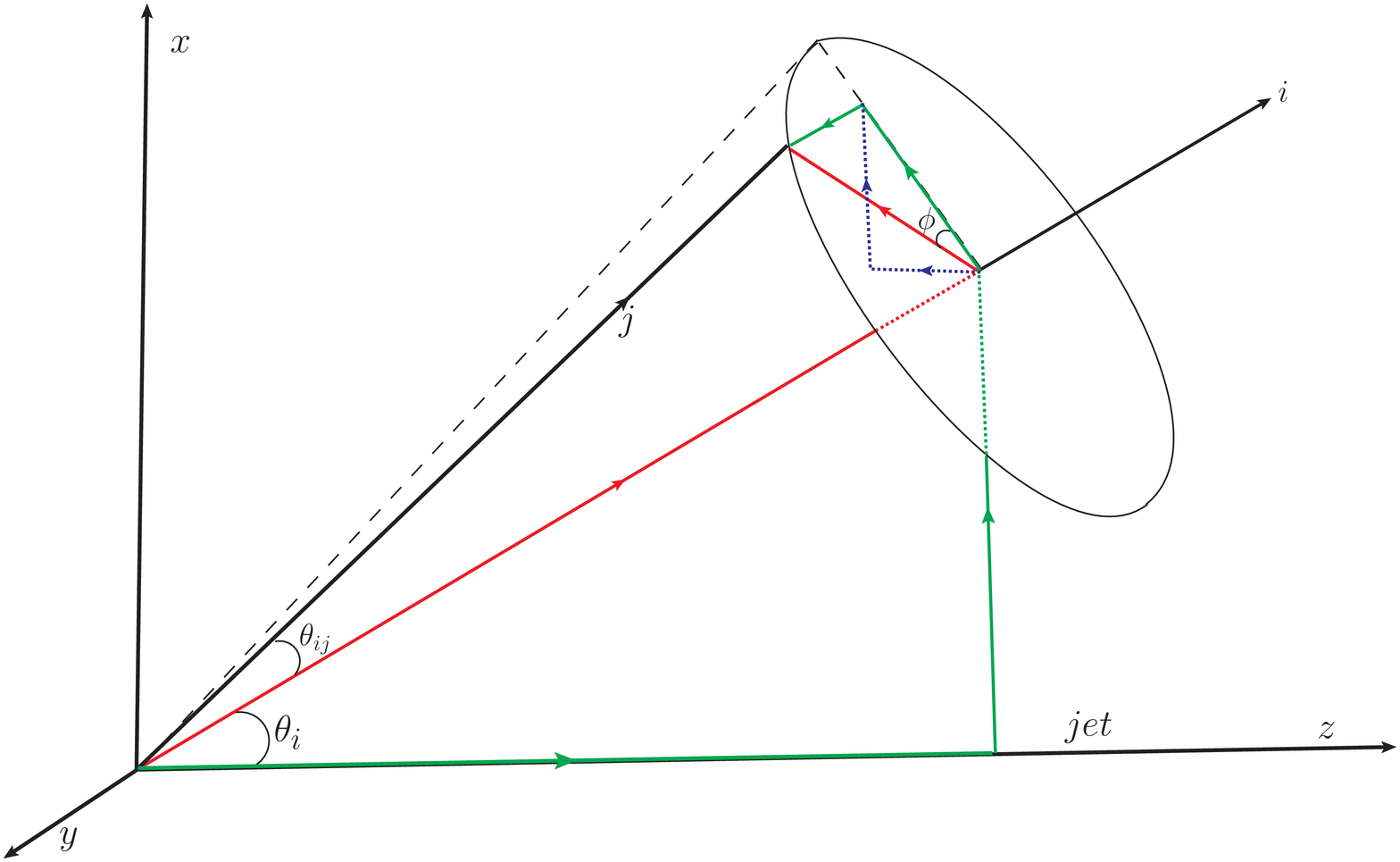}
   \caption{(Color online) The geometry for a parton $i$ (inside a jet) radiating a parton $j$, where $\theta_i$ and $\theta_j$ are the angles of the parton $i$ and the parton $j$ with repect to the direction of jet axis ($z$ axis in the figure), and $\theta_{ij}$ is the relative angle of the parton $j$ with respect to the propagation direction of the parton $i$. }
  \label{fig:geometry}
\end{figure}

To solve the above differential equation, the splitting rates must be supplied. 
In this work, we use results from higher-twist energy loss formalisms\cite{Wang:2001ifa, Majumder:2009ge} in which the medium-induced gluon radiation spectrum takes the following form: 
\begin{eqnarray}
\label{eq:Gmed}
\frac{d\Gamma_g(\omega, k_\perp^2|E)} {d\omega dk_\perp^2 dt} = \frac{2 \alpha_s}{\pi} \frac{ x P(x) \hat{q}_g(t) }{\omega k_\perp^4} \sin^2 \left(\frac{t - t_i}{2\tau_f}\right).
\end{eqnarray}
In the above equation, $P(x=\omega/E)$ is the vacuum splitting function for a given process, with $\omega$ the energy of the radiated parton and $E$ the energy of the parent parton, and $\hat{q}_g$ is the gluon transverse momentum broadening rate, $\tau_f = 2Ex(1-x)/k_\perp^2$ denotes the formation time for the radiation, with $k_\perp$ the transverse momentum of the radiated parton with respect to the propagation direction of the parent parton, and $t_i$ is the production time of the parent parton.
To apply the above radiation spectrum to the jet evolution equation, one has to calculate the relative transverse momenta in the radiation process $i \to j$, where the parton $i$ carries energy $\omega_i$ and the transverse momentum $k_{i\perp}$, and the parton $j$ carries the energy $\omega_j$ and the transverse momentum $k_{j \perp}$, i.e.,  
\begin{eqnarray}
\label{eq:Gmed-Jacobian}
\frac{d\tilde{\Gamma}_{i\rightarrow j}(\omega_j, k_{j\perp}^2|\omega_i, k_{i\perp}^2)}{d\omega_j dk_{j\perp}^2 dt} = \mathcal{J} \frac{d\Gamma_{i \rightarrow j}(\omega_j, k_{ij \perp}^2| \omega_i)} {d\omega_j dk_{ij\perp}^2 dt},
\end{eqnarray}
where $\mathcal{J} = \left|\frac{\partial k_{ij\perp}^2}{\partial k_{j\perp}^2}\right|$ is the Jacobian.
The relation between $k_{ij\perp}^2$ and $k_{j\perp}^2$ can be obtained using Fig. \ref{fig:geometry} which shows the geometry for a parton $i$ (within a jet) radiating another parton $j$. 
The angles of the parton $i$ and the parton $j$ with repect to the jet axis (i.e., $z$ direction in the figure) are denoted as $\theta_i$ and $\theta_j$, and the relative angle between the parton $i$ and the parton $j$ is denoted as $\theta_{ij}$. 

From Fig.~\ref{fig:geometry}, we can write down the following geometric relation between different transvese momenta and angles:
 \begin{eqnarray}
\label{eq:theta_ij}
k^2_{j\perp}&=&k^2_{jx}+k^2_{jy} \nonumber\\
&=&\omega^2_j\left[(\cos\theta_{ij}\sin\theta_i+\sin\theta_{ij}\cos\phi\cos\theta_i)^2
\right.
\nonumber\\&& 
\left. 
+(\sin\theta_{ij}\sin\phi)^2\right].
\end{eqnarray}
We may perform the average over the azimuthal angle $\phi$ (see Fig. \ref{fig:geometry}) for the above equation and obtain: 
\begin{eqnarray}
\label{eq:ave_kj}
\langle k^2_{j\perp}\rangle 
&=& \omega^2_j\left[\cos^2\theta_{ij}\sin^2\theta_i + \frac{1}{2} \sin^2\theta_{ij}(1+\cos^2\theta_i)\right]
\nonumber\\
&=& \omega^2_j\left(\frac{k_{i\perp}^2}{\omega_i} + \frac{k_{ij\perp}^2}{\omega_j} - \frac{3}{2} \frac{k_{i\perp}^2}{\omega_i} \frac{k_{ij\perp}^2}{\omega_j} \right).
\end{eqnarray}
From the above relation, one may calculate the Jacobian $\mathcal{J}$ in Eq. (\ref{eq:Gmed-Jacobian}). 
We can see that for collinear radiations, i.e., the angles $\theta_{ij}$ and $\theta_i$ are very small, the following simple approximate relations can be obtained:
\begin{eqnarray}
\label{eq:ave_kj2}
\langle \theta_j^2 \rangle \approx \langle \frac{k_{j\perp}^2}{\omega_j^2} \rangle \approx \theta_i^2+\theta_{ij}^2 \approx \frac{k_{i\perp}^2}{\omega_i^2} + \frac{k_{ij\perp}^2}{\omega_j^2}.
\end{eqnarray}
We note that for the results presented in this article, the exact relation in Eq.~(\ref{eq:ave_kj}) instead of Eq.~(\ref{eq:ave_kj2}) is used. 

In our current study, we consider the three-dimensional momentum evolutions of the quarks (plus anti-quark) and gluons contained in the full jet, therefore, the explicit expressions for the coupled jet evolution equations read as follows:
\begin{eqnarray}
\label{eq:dG/dt_q}
\!\!&&\!\!\frac{d}{dt}f_q(\omega_q, k_{q\perp}^2, t) = \left(\hat{e}_q \frac{\partial}{\partial \omega_q}
  + \frac{1}{4} \hat{q}_q {\nabla_{k_{q\perp}}^2 }\right) f_q(\omega_q, k_{q\perp}^2, t) \ \ \ \nonumber\\
\!\!&&\!\!+ \int d\omega_q'dk_{q\perp}'^2 \frac{d\tilde{\Gamma}_{q\rightarrow qg}(\omega_q, k_{q\perp}^2|\omega_q', k_{q\perp}'^2)}{d\omega_q dk_{q\perp}^2}f_q(\omega_q', k_{q\perp}'^2, t)\nonumber\\
\!\!&&\!\!+n_f \int d\omega_g'dk_{g\perp}'^2 \frac{d\tilde{\Gamma}_{g\rightarrow q\bar{q}}(\omega_q, k_{q\perp}^2|\omega_g', k_{g\perp}'^2)}{d\omega_q dk_{q\perp}^2}f_g(\omega_g', k_{g\perp}'^2, t)\nonumber\\
\!\!&&\!\!- \int d\omega_q'dk_{q\perp}'^2 \frac{\tilde{\Gamma}_{q\rightarrow qg}(\omega_q', k_{q\perp}'^2|\omega_q, k_{q\perp}^2)}{d\omega_q' dk_{q\perp}'^2}f_q(\omega_q, k_{q\perp}^2, t);
\end{eqnarray}
\begin{eqnarray}
\label{eq:dG/dt_g}
\!\!&&\!\!\frac{d}{dt}f_g(\omega_g, k_{g\perp}^2, t) = \left(\hat{e}_g \frac{\partial}{\partial \omega_g}
  + \frac{1}{4} \hat{q}_g {\nabla_{k_{g\perp}}^2 }\right) f_g(\omega_g, k_{g\perp}^2, t)\ \ \ \nonumber\\
\!\!&&\!\!+ \int d\omega_g'dk_{g\perp}'^2 \frac{d\tilde{\Gamma}_{g\rightarrow gg}(\omega_g, k_{g\perp}^2|\omega_g', k_{g\perp}'^2)}{d\omega_g dk_{g\perp}^2}f_g(\omega_g', k_{g\perp}'^2, t)\nonumber\\
\!\!&&\!\!+ \int d\omega_q'dk_{q\perp}'^2 \frac{d\tilde{\Gamma}_{q\rightarrow gq}(\omega_g, k_{g\perp}^2|\omega_q', k_{q\perp}'^2)}{d\omega_g dk_{g\perp}^2}f_q(\omega_q', k_{q\perp}'^2, t)\nonumber\\
\!\!&&\!\!- \frac{1}{2}\int d\omega_g'dk_{g\perp}'^2 \frac{d\tilde{\Gamma}_{g\rightarrow gg}(\omega_g', k_{g\perp}'^2|\omega_g, k_{g\perp}^2)}{d\omega_g' dk_{g\perp}'^2}f_g(\omega_g, k_{g\perp}^2, t)\,\nonumber\\
\!\!&&\!\!- \frac{n_f}{2} \int d\omega_q'dk_{q\perp}'^2 \frac{d\tilde{\Gamma}_{g\rightarrow q\bar{q}}(\omega_q', k_{q\perp}'^2|\omega_g, k_{g\perp}^2)}{d\omega_q' dk_{q\perp}'^2}f_g(\omega_g, k_{g\perp}^2, t).
\end{eqnarray}
The transport coefficients for quarks and gluons are related by a color factor:  $\hat{q}_g/\hat{q}_q = C_A/C_F =  \hat{e}_q/\hat{e}_q$.  

\begin{center}
\begin{figure}[tbhps]
  \centering
     \includegraphics[width=1.0\linewidth]{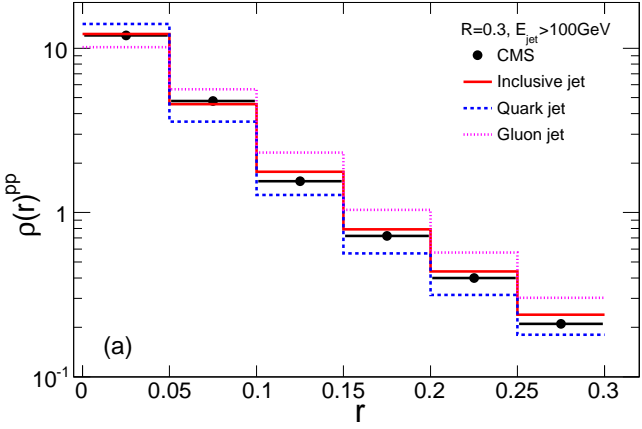}
  \caption{(Color online) The jet shape function for $p$+$p$ collisions at 2.76~TeV with $p_T>100$~GeV, compared to CMS data \cite{Chatrchyan:2013kwa}.}
  \label{fig:tho}
\end{figure}
\end{center}

In order to solve the above coupled differential equations, Eq.~(\ref{eq:dG/dt_q}) and Eq.~(\ref{eq:dG/dt_g}), the initial conditions, i.e., the three-dimensional momentum distributions of the quarks (plus anti-quarks) and gluons before entering the medium, must be provided. 
We generate the initial conditions using PYTHIA simulation \cite{Sjostrand:2007gs}. 
The parameters in PYTHIA have been tuned such that jet shape function in $p$+$p$ collisions at 2.76~TeV at the LHC \cite{Chatrchyan:2013kwa} can be described, as shown in Fig. \ref{fig:tho} (see Sec. \ref{sec:results} for the definition of jet shape function and more discussions). 
The effect of the medium on the transport of the full jet, in terms of the three-dimensional momentum distributions of the shower partons, can be investigated using the above coupled evolution equations. 
As has been mentioned, we include the efffect that the shower partons can also experienced inelastic radiative processes after they are formed. 
This is implemented in our calculation by a restriction that the medium-induced radiation at time $t$ is only allowed for the shower partons with the formation time $\tau_f<t$. 
We also impose a minimum cutoff $\omega_{\rm min}$~GeV for the radiation (set as $2$~GeV in our current calculation) to take into account the balance effect from the medium-induced absorption.
The same cutoff energy is used for the shower partons during their evolution assuming that the patrons with energy lower than $\omega_{\rm min}$ are thermalized and then treated as a part of the medium.

\section{Numerical results for various full jet observables}
\label{sec:results}

In this section, we present our numerical results for the nuclear modifications of various full jet observables in Pb+Pb collisions at 2.76~ATeV at the LHC, including the nuclear modification factor for single inclusive jet spectra, the momentum imbalance distributions for dijet and photon-jet pairs, and the nuclear modification of jet shape function (at partonic level). 
The dynamical evolution and space-time profiles such as the local temperature and flow velocity of the hot and dense QGP medium created in Pb+Pb collisions are simulated using the (2+1)-dimensional viscous hydrodynamics model (VISH2+1) developed by The Ohio State University group \cite{Song:2007ux, Chen:2010te, Qiu:2011hf}.
The coupled differential transport equations, Eq.~(\ref{eq:dG/dt_q}) and Eq.~(\ref{eq:dG/dt_g}), are solved numerically to study the evolution and modification of the full jet shower in the hot and dense nuclear medium.  
One of the key quantities in determining the medium effect on the full jet evolution are the transport coefficients $\hat{e}$ and $\hat{q}$, which not only control the effect of elastic collisions on the shower partons, but also govern the amount of medium-induced radiation. 
In this work, we relate $\hat{q}$ to the local temperature and flow velocity of the QGP medium by dimensional argument as follows:
\begin{equation}
\label{eq:qhat}
\hat{q} (\tau,\vec{r}) = \hat{q}_0 \cdot \frac{T^3(\tau,\vec{}r)}{T_{0}^3(\tau_{0},\vec{0})} \cdot \frac{p\cdot u(\tau, \vec{r})}{p_0}. 
\end{equation}
where $p^\mu$ is the four-momentum of the propagating parton, $u_\mu$ is the local four-velocity of the medium flow, and the multiplicative factor ${p\cdot u}/{p_0}$ is to account for the flow effect on the jet modification, i.e., the effective value of $\hat{q}$ with the presence of the medium flow is different from that in a static medium \cite{Baier:2006pt}. 
In the above equation, $\hat q_{0}$ denotes the value of the transport parameter $\hat{q}$ for a quark jet at the center of the QGP at the initial time $\tau_{0}=0.6~{\rm fm}/{\rm c}$ in most central (0-10\%) collisions. 
From the value of $\hat{q}$, we obtain the collision energy loss rate $\hat{e}$ via the relation $\hat{q} = 4T \hat{e}$, with the assumption that the medium is close to local thermal equilibrium and the fluctuation-dissipation theorem can be applied \cite{Qin:2009gw}.
With the above setup, there is only one parameter $\hat{q}_0$ in our calculation, which we may tune to describe one set of jet quenching observable. 

We first fix our model parameter $\hat{q}_0$ by comparing to the experimental measurements of the nuclear modification factor $R_{AA}$ for single inclusive jet spectra, which is defined as follows:
\begin{equation}
\label{eq:raa}
R_{AA}= \frac{1}{\langle N_{coll}\rangle}\frac{d^2N_{AA}/d\eta dp_T}{d^2N_{pp}/d\eta dp_T},
\end{equation}
where $\langle N_{coll}\rangle$ is the event-averaged number of binary nucleon-nucleon collisions at a given centrality class. 
In our study, the number of the binary nucleon-nucleon collisions as well as their spatial distribution in the transverse plane in Pb+Pb collisions are calculated using the Glauber model \cite{Miller:2007ri}; the later is used to simulate the distribution of initial jet production points. 
The full jet spectrum in $p$+$p$ collisions is obtained via PYTHIA simulation \cite{Sjostrand:2007gs} combined with FASTJET package for full jet reconstruction \cite{Cacciari:2011ma}. 
After fixing the initial production point and the propagation direction for a full jet, its propagation and evolution in the hot and dense medium in terms of the quarks' and gluons' three-dimensional momentum distribution is described according to Eq.~(\ref{eq:dG/dt_q}) and Eq.~(\ref{eq:dG/dt_g}). 
Here we only include the interaction of jets and medium in the partonic phase, and the medium effect on jets in hadronic phase is usually small and neglected in the current study. 
Thus, when the local medium temperature drops below $T_{c}=160$~MeV, we turn off the jet-medium interaction.
After the jet exits the partonic medium, we use the three-dimensional momentum distribution of quarks and gluons contained in the full jet to calculate various quantities for the quenched full jet, and compare to initial unquenched jets. 

\begin{center}
\begin{figure}[tbhps]
  \centering
     \includegraphics[width=1.0\linewidth]{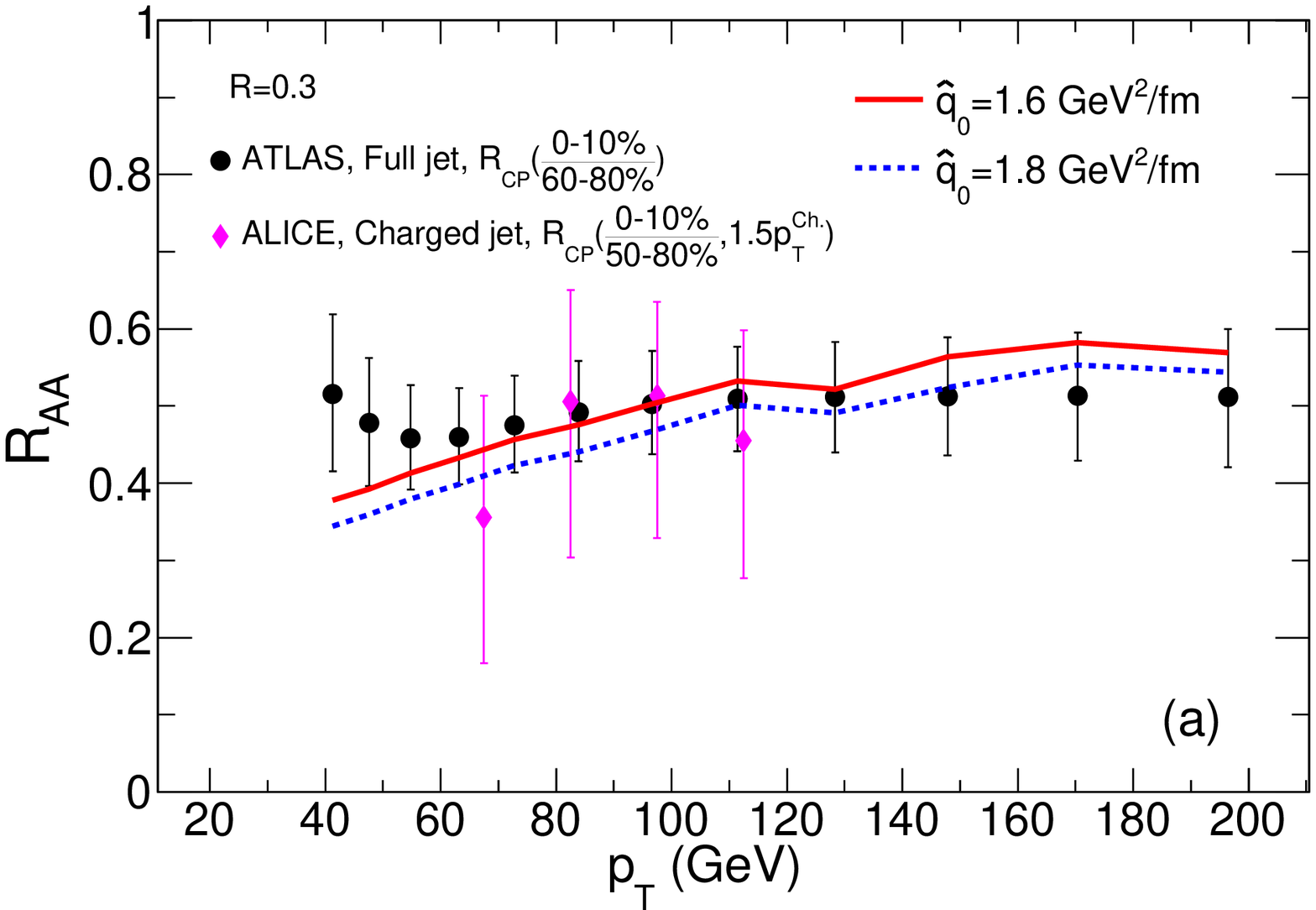}
     \includegraphics[width=1.0\linewidth]{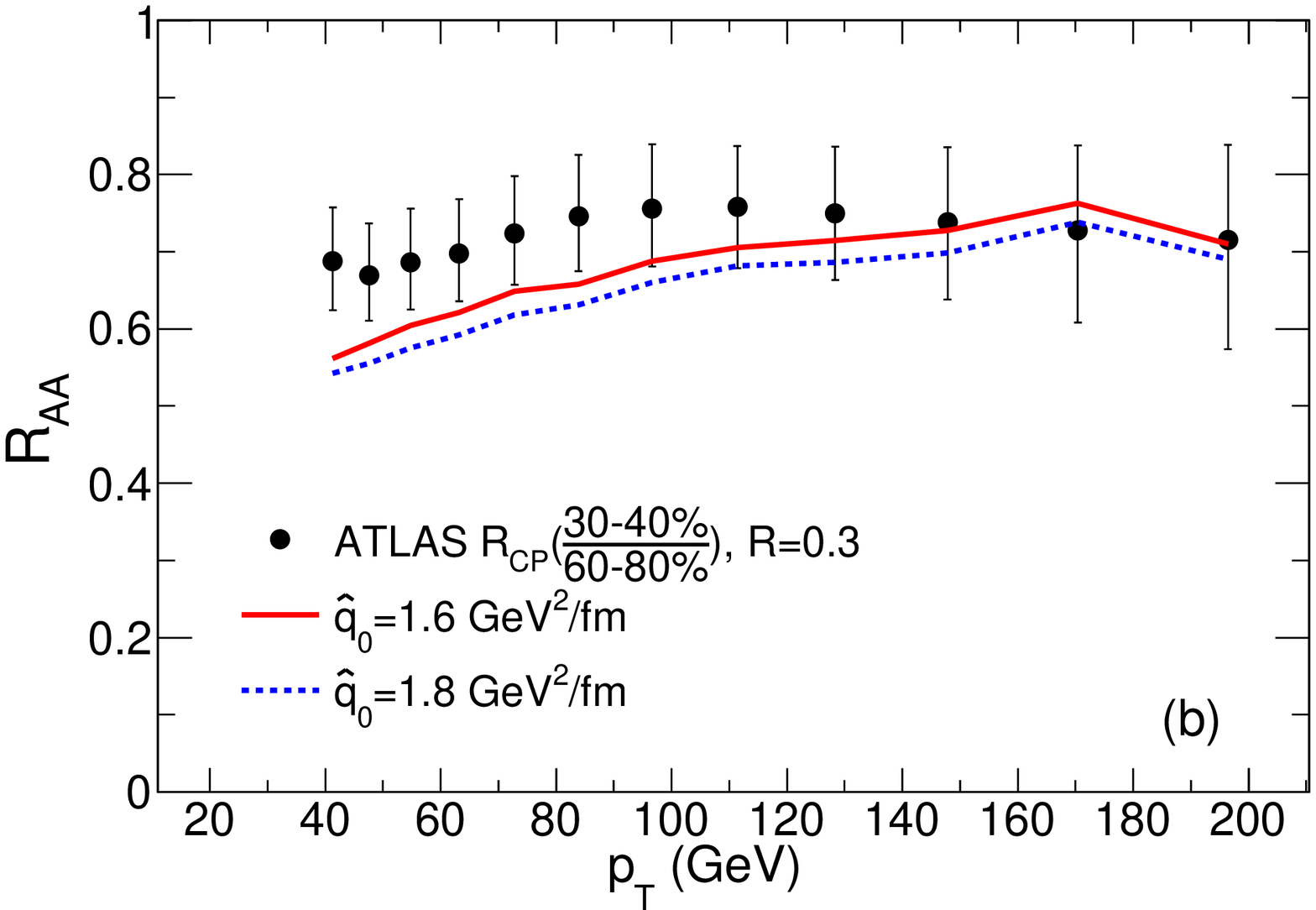}
  \caption{(Color online) The nuclear modification factor $R_{AA}$ for single inclusive full jet production in (a) most central $0$-$10$\%  and (b) mid-central $30$-$40$\% Pb+Pb collisions at 2.76~ATeV, compared to the full jet $R_{CP}$ measured by the ATLTAS and ALICE Collaborations \cite{Aad:2012vca,Abelev:2013kqa}.  }
  \label{fig:raa}
\end{figure}
\end{center}

In Fig.~\ref{fig:raa}, we show our numerical results for the nuclear modification factor $R_{AA}$ of single inclusive full jet spectra in Pb+Pb collisions at 2.76~ATeV at the LHC: the panel (a) for most central $0$-$10$\% collisions, and the panel (b) for mid-central $30$-$40$\% collisions, compared to the full jet $R_{CP}$ measured by the ATLAS and ALICE Collaborations \cite{Aad:2012vca, Abelev:2013kqa}. 
In our calculation, we impose a Gaussian smearing for the jet energies and jet spectra in both $p$+$p$ and Pb+Pb collisions to take into account the hadronization effect and the reconstruction efficiency in the expreiment measurements.
As the reconstruction efficiency varies in different experiments, we fix the smearing factor by comparing to full jet spectra measured in the reference $p$+$p$ collisions \cite{Abelev:2013fn}. 
From the figure, we can see that our model calculation can give a reasonable description of the nuclear modification factor $R_{AA}$ for single inclusive full jets in both most central and mid-central Pb+Pb collisions at 2.76~ATeV.
We also note the deviation in the smaller $p_T$ regime where there exist larger uncertainties in the experimental measurements as well. 
From fitting to the single inclusive full jet $R_{AA}$, we obtain the value for our model parameter as $\hat{q}_0 \approx 1.7$~GeV$^2/$fm for a quark jet at the center of the medium at the initial time $\tau_0$ in most central 0-10\% Pb+Pb collisions at 2.76~ATeV. 
This value is consistent with the value obtained by JET Collaboration \cite{Burke:2013yra}, in which the jet transport parameter was extracted by comparing five different jet energy loss model calculations to the experimental measurements of single inclusive hadron $R_{AA}$ in most central Au+Au collisions at $200$~AGeV at RHIC and in most central Pb+Pb collisions at $2.76$~ATeV at the LHC.
The calculation of single inclusive full jet $R_{AA}$ presented in Fig.~\ref{fig:raa} provides the baseline to study other full jet observables and investigate in more details the roles of different jet-medium interaction mechanisms for the medium modification of full jets. 

\begin{center}
\begin{figure}[tbhps]
  \centering
     \includegraphics[width=1.0\linewidth]{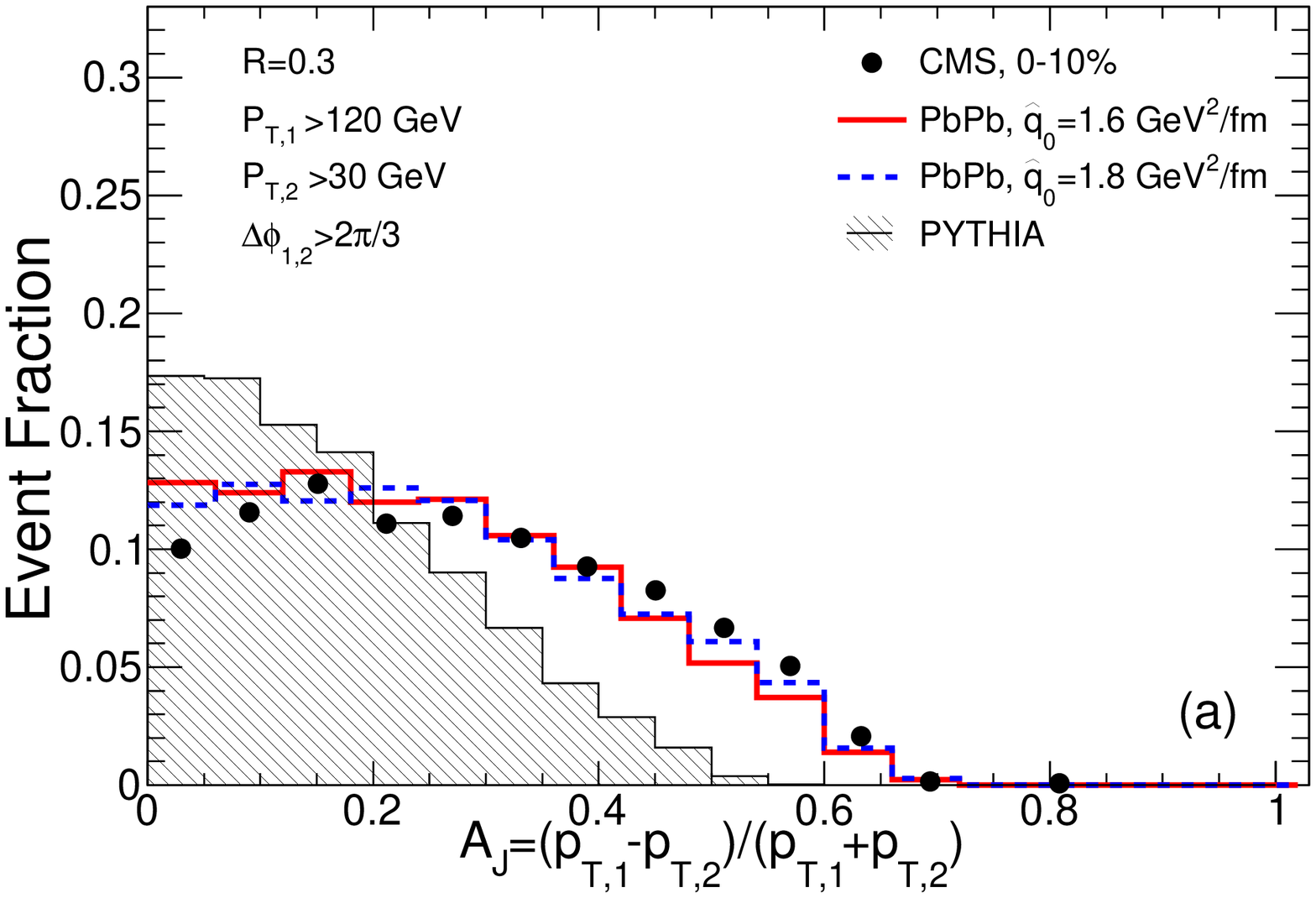}
     \includegraphics[width=1.0\linewidth]{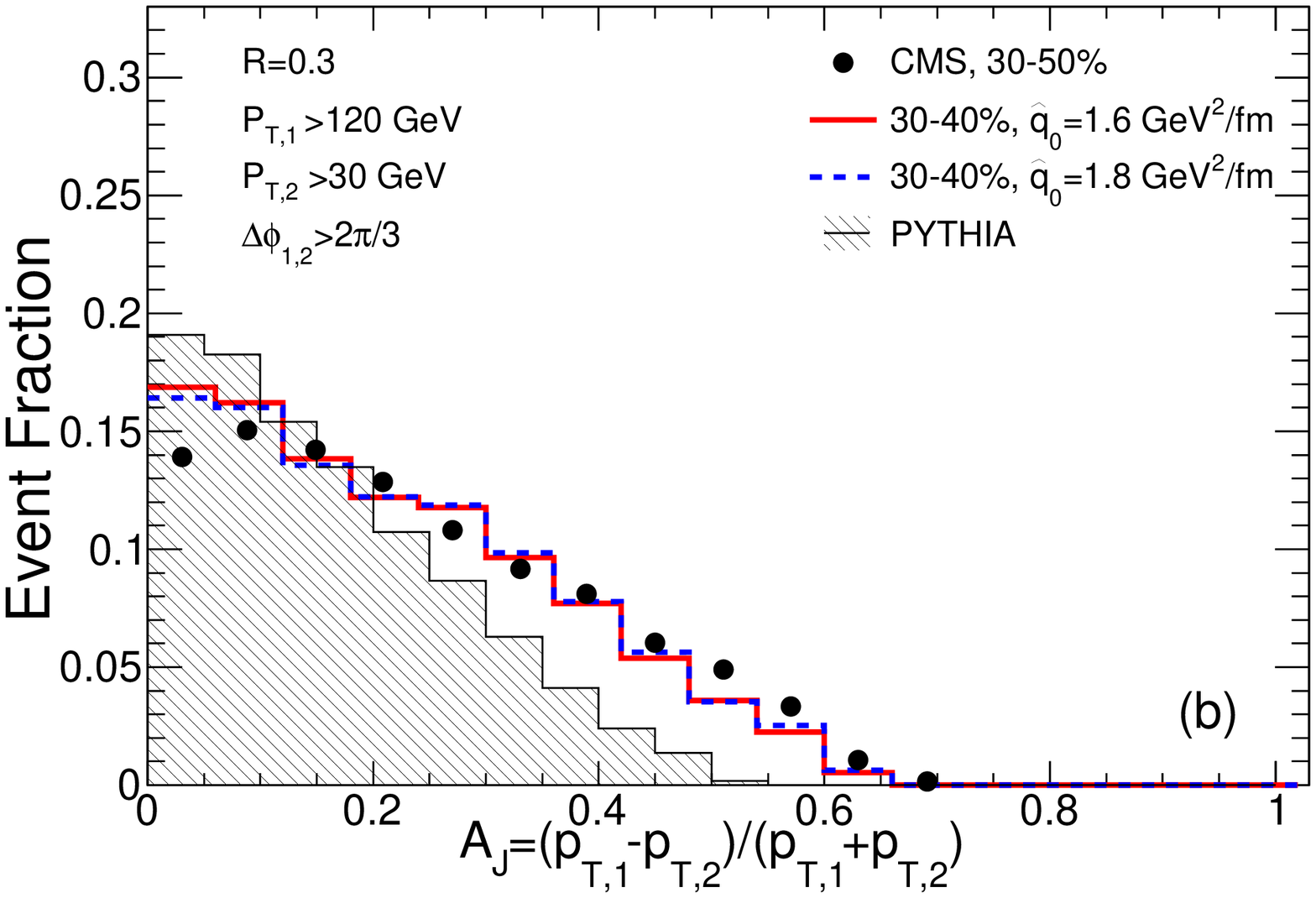}
  \caption{(Color online) The event distribution of dijet momentum imbalance $A_J$ for (a) most central $0$-$10$\% and (b) mid-central $30$-$40$\% Pb+Pb collisions at 2.76~ATeV, compared to the data from CMS Collaboration \cite{Chatrchyan:2012nia}. }
  \label{fig:dijet}
\end{figure}
\end{center}

\begin{center}
\begin{figure}[tbhps]
  \centering
     \includegraphics[width=1.0\linewidth]{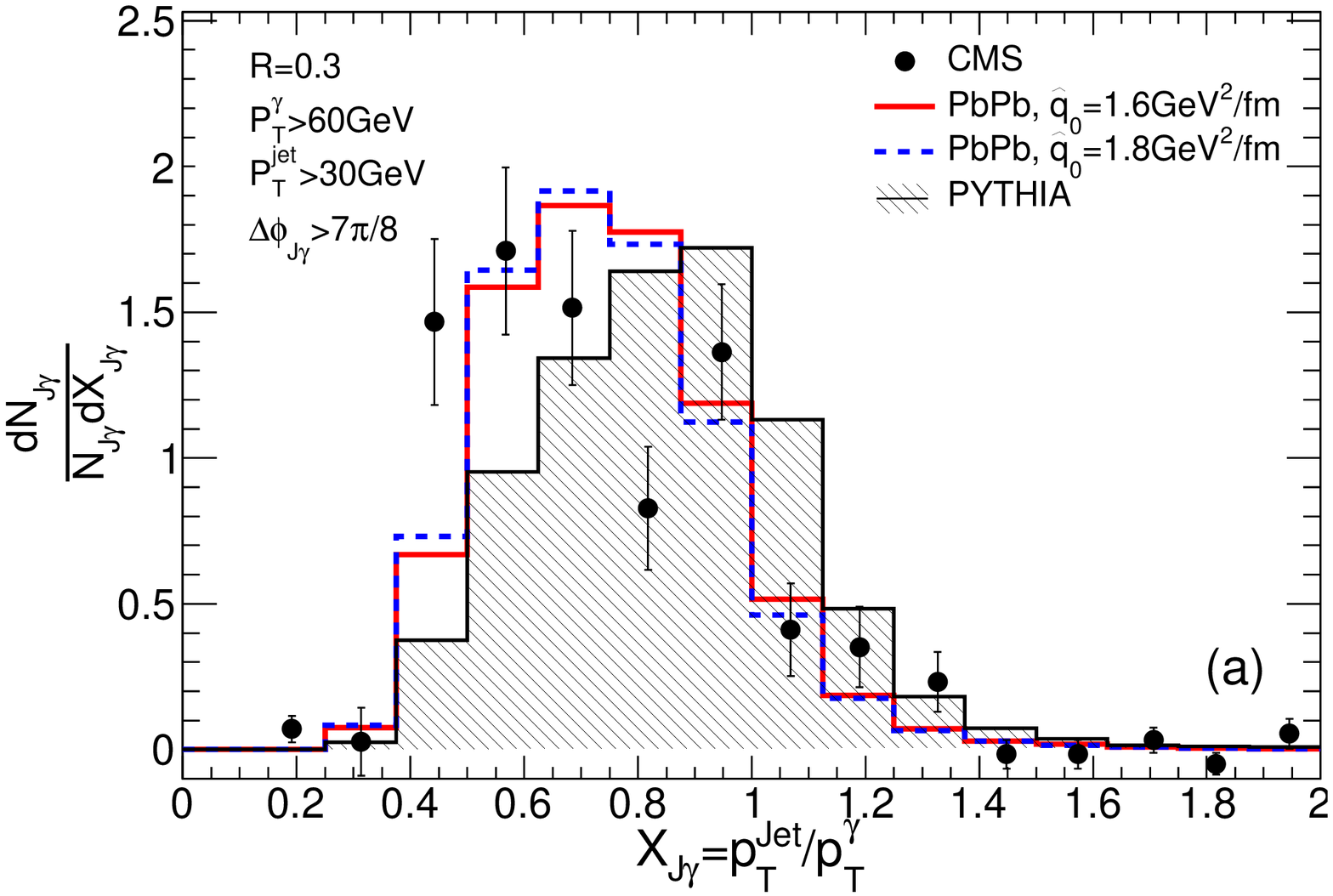}
     \includegraphics[width=1.0\linewidth]{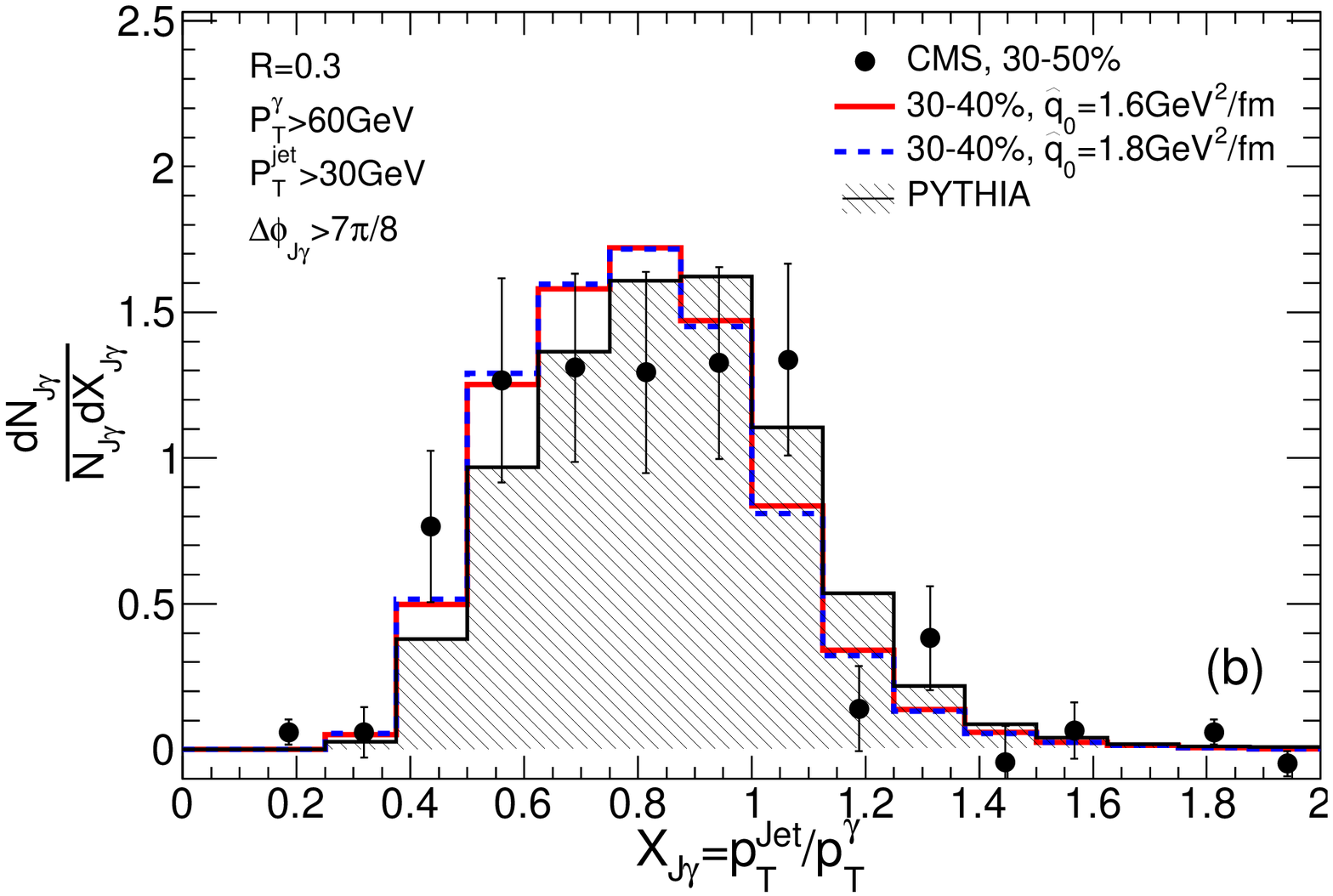}
  \caption{(Color online) The event distribution of photon-jet momentum imbalance variable $x_{J\gamma}$ in (a) most central $0$-$10$\%  and (b) mid-central $30$-$40$\%  Pb+Pb collisions at 2.76~ATeV, comared to the data from CMS Collaboration \cite{Chatrchyan:2012gt}. }
  \label{fig:gamma-jet}
\end{figure}
\end{center}

We now present our numerical results for the nuclear modification of dijet momentum imbalance $A_J$ distribution in Pb+Pb collisions at 2.76~ATeV at the LHC, where the momentum imbalance factor $A_J$ is defined as:
\begin{eqnarray}
A_J = \frac{p_{T, 1} - p_{T, 2}}{p_{T, 1} + p_{T, 2}},
\end{eqnarray}
with $p_{T,1}$ and $p_{T, 2}$ the transverse momenta of the leading and subleading jets. 
The nuclear modification of dijet momentum imbalance distribution was one of the first measurements of full jets in Pb+Pb collisions after the launch of the LHC \cite{Aad:2010bu}, which showed a significant change of the dijet $A_J$ distribution in central Pb+Pb collisions compared to that in the reference $p$+$p$ collisions. 
Various studies have shown that the measured modification results for dijet $A_J$ distribution can be explained quite well by the energy loss experienced by the full jets when they propagate through and interact with the hot and dense nuclear medium produced in relativistic heavy-ion collisions \cite{Qin:2010mn,CasalderreySolana:2010eh,Lokhtin:2011qq,Young:2011qx,He:2011pd,Renk:2012cx,Ma:2013pha,Senzel:2013dta,Chien:2014nsa,Milhano:2015mng}. 
In Fig.~\ref{fig:dijet}, we show our calculation for dijet momentum imbalance $A_J$ distribution in most central $0$-$10$\% and mid-central ($30$-$40$\%) Pb+Pb collisions at 2.76~ATeV, compared with the measurements from CMS Collaboration \cite{Chatrchyan:2012nia}. 
In this calculation, the smearing factor for the jet energies is tuned such that the experimental `PYTHIA+HYDJET' simulation results can be described. 
As we can see, using the same values for the jet transport parameter $\hat{q}_{0}$ as in the single inclusive full jet $R_{AA}$ calculation, our model can give a good description of the nuclear modification of dijet momentum imbalance distribution in Pb+Pb collisions at 2.76~ATeV at the LHC. 

Photon-tagged jets have been proposed to be one of the golden channels for studying jet energy loss and jet quenching in relativisitic heavy-ion collisions \cite{Wang:1996yh,Zhang:2009rn} due to the fact that the photon transverse momentum is a good reference for that of the away-side partonic jet at the initial production time (We note that this is only true at the leading order). 
The momentum imbalance distribution of photon-jet pairs has also been measured in Pb+Pb collisions at 2.76~ATeV at the LHC \cite{Chatrchyan:2012gt}, and the result is quite similar to dijet pairs. 
Various phenomenological calculations have also been performed for the nuclear modification of photon-jet momentum imbalance distribution \cite{Dai:2012am,Qin:2012gp,Wang:2013cia}.
In Fig.~\ref{fig:gamma-jet}, we show our current calculation of the event distribution for the momentum balance variable $x_{J\gamma} = p_{T,J}/p_{T,\gamma}$, defined as the ratio of the jet transverse momentum $p_{T, J}$ to the trigger photon transvere momentum $p_{T, \gamma}$, in most central $0$-$10$\% and mid-central $30$-$40$\% Pb+Pb collisions at 2.76~ATeV, compared with the measurements from CMS Collaboration \cite{Chatrchyan:2012gt}. 
Both our calculation and the experimental data show that the event distribution of $x_{J\gamma}$ moves towards smaller value of $x_{J\gamma}$ due to the energy loss of the away-side jet during the proagation through the hot and dense medium produced in the collisions. 

\begin{center}
\begin{figure}[tbhps]
  \centering
     \includegraphics[width=1.0\linewidth]{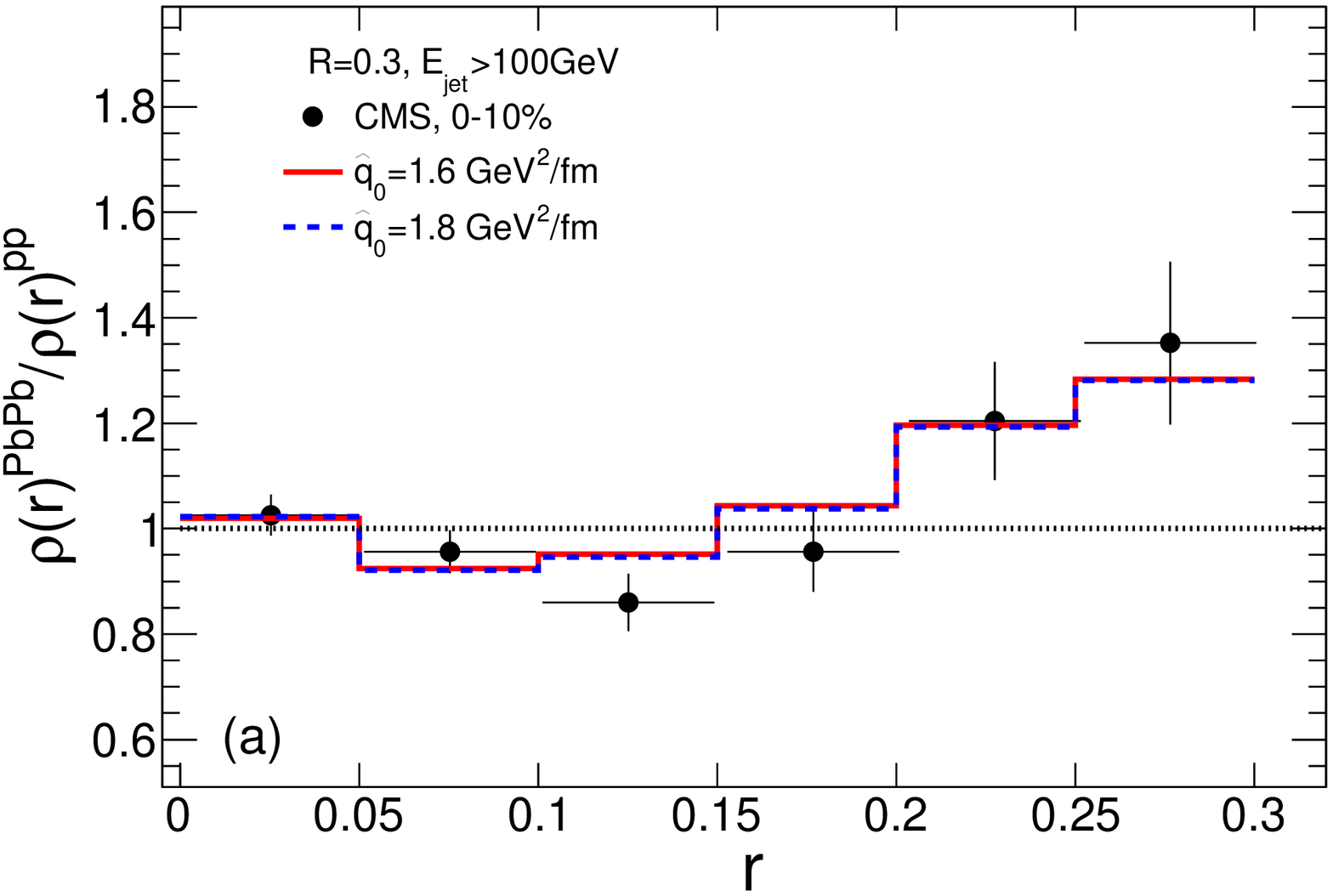}
     \includegraphics[width=1.0\linewidth]{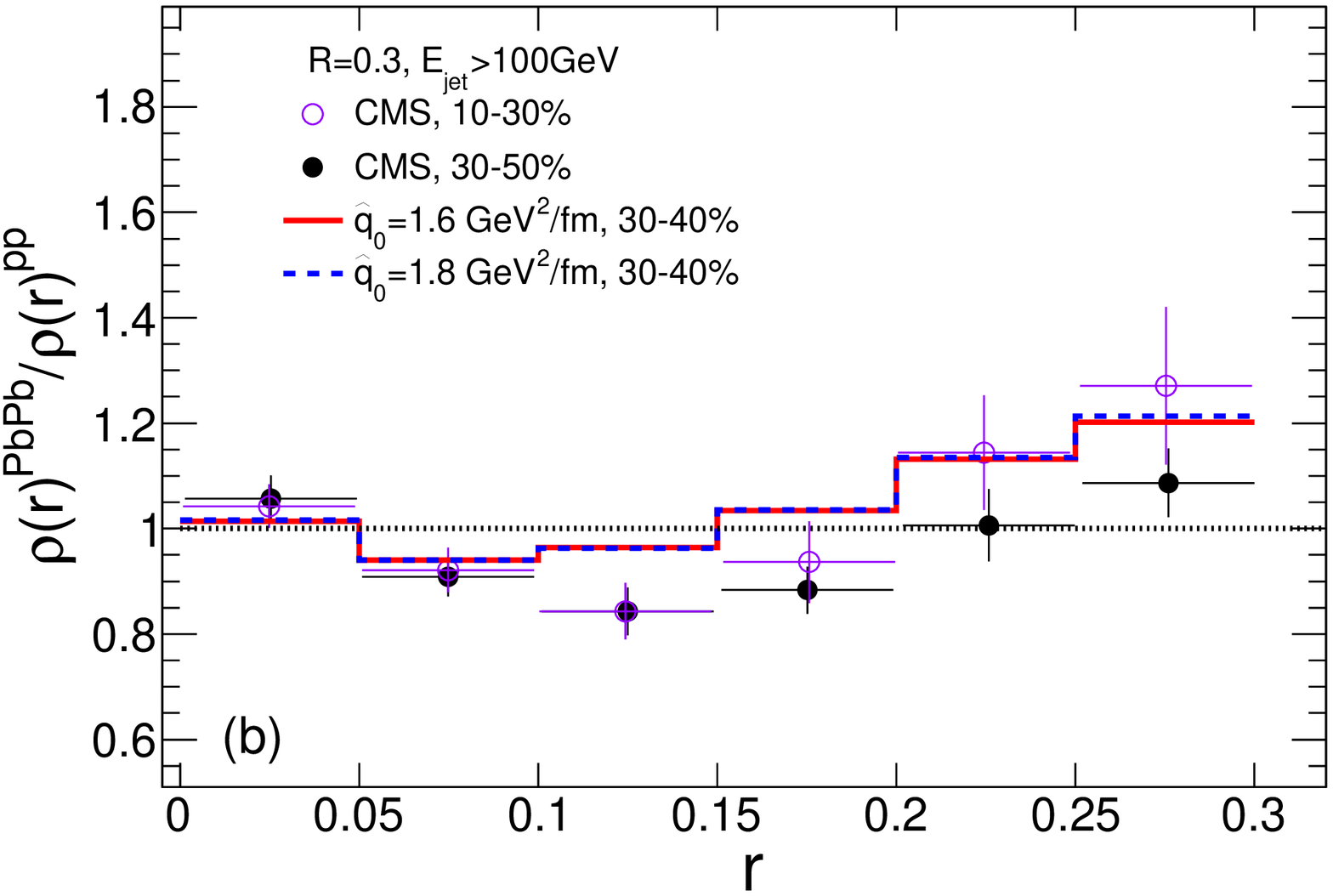}
  \caption{(Color online) The nuclear modification factor $R_{AA}^{\rho}(r) = \rho_{AA}(r)/\rho_{pp}(r)$ for the differential jet shape function in (a) most central $0$-$10$\% and (b) mid-central $30$-$40$\% Pb+Pb collisions at 2.76~ATeV. The experimental data are taken from CMS Collaboration \cite{Chatrchyan:2013kwa}. }
  \label{fig:rtho}
\end{figure}
\end{center}

Up to now, we have presented the results for the nuclear modifications of single inclusive jet spectra and the momentum imbalance of dijet and photon-jet pairs in Pb+Pb collisions at 2.76~ATeV. 
As is known, these observables are only sensitive to the total energy (loss) of the full jets. 
One of the advantages of full jets is that one can look at their internal strucutres and study how the distribution of the energy and momentum within the jet cone are modified due to the interaction with the medium. 
One popular quantity is jet shape function which describes the energy distribution along the radial/transverse direction of the full jet. 
In the experiments, the (differential) jet shape function is defined as follows:
\begin{eqnarray}
\label{eq:rho_r}
\rho_{\rm jet}(r) = \sum_{i} \frac{ p_T^i}{p_T^{\rm jet}} \frac{\theta[r_i - (r-\frac{1}{2}\delta r)] \theta[(r+\frac{1}{2}\delta r)-r_i]}{\delta r},
\end{eqnarray}
where $r_i = \sqrt{(\eta_i - \eta_{\rm jet})^2 + (\phi_i - \phi_{\rm jet})^2}$, $\delta r$ is the size of the bin, and the sum over $i$ runs over all constituents of the full jets with a given cone size.
In the real analysis, one also performs the average over all the jets that satisfy the required condition: $\rho(r) = \frac{1}{N_{\rm jet}} \sum_{\rm jet} \rho_{\rm jet}(r)$.
It is easy to see that the above-defined differential jet shape function is normalized to unity, $\int \rho(r) dr = 1$.

In Fig.~\ref{fig:rtho}, we show the nuclear modification factor $R_{AA}^{\rho}(r) = \rho_{AA}(r)/\rho_{pp}(r)$ for the single inclusive jet shape function $\rho(r)$ in most central $0$-$10$\% (a) and mid-central $30$-$40$\% (b) Pb+Pb collisions at 2.76~ATeV at the LHC, compared with the data from CMS Collaboration \cite{Chatrchyan:2013kwa}. 
In this figure, we put the same jet energy cut as CMS Collaboration: $p_T > 100$~GeV. 
One can see that our calculation shows the same modification pattern as the measurements from the CMS Collaboration, i.e., an enhancement at larger values of $r$ and a depletion at intermediate values of $r$. 
The enhancement at larger values of $r$ can be understood as the combinational effect of medium-induced radiation and transverse momentum broadening. 
Since the differential jet function $\rho(r)$ is normalized to unity, the enhancement at larger $r$ must be compensated by some depletion at smaller values of $r$. 
This also implies that the nuclear modification of $\rho(r)$ at large values of $r$ is very sensitive to the modification at smaller values of $r$, since $\rho(r)$ is a deep-falling function of $r$. 
We also note that the modification at the very small values of $r$ is very small. 
This is because for high energy jets, the inner hard core of the full jets is very difficult to be modified by the medium in contrast to the outer soft part of the full jets. 
Below, we will investigate in more details the jet energy and flavor dependence for the nuclear modification of jet shape function.

\begin{center}
\begin{figure}[tbhps]
  \centering
     \includegraphics[width=1.0\linewidth]{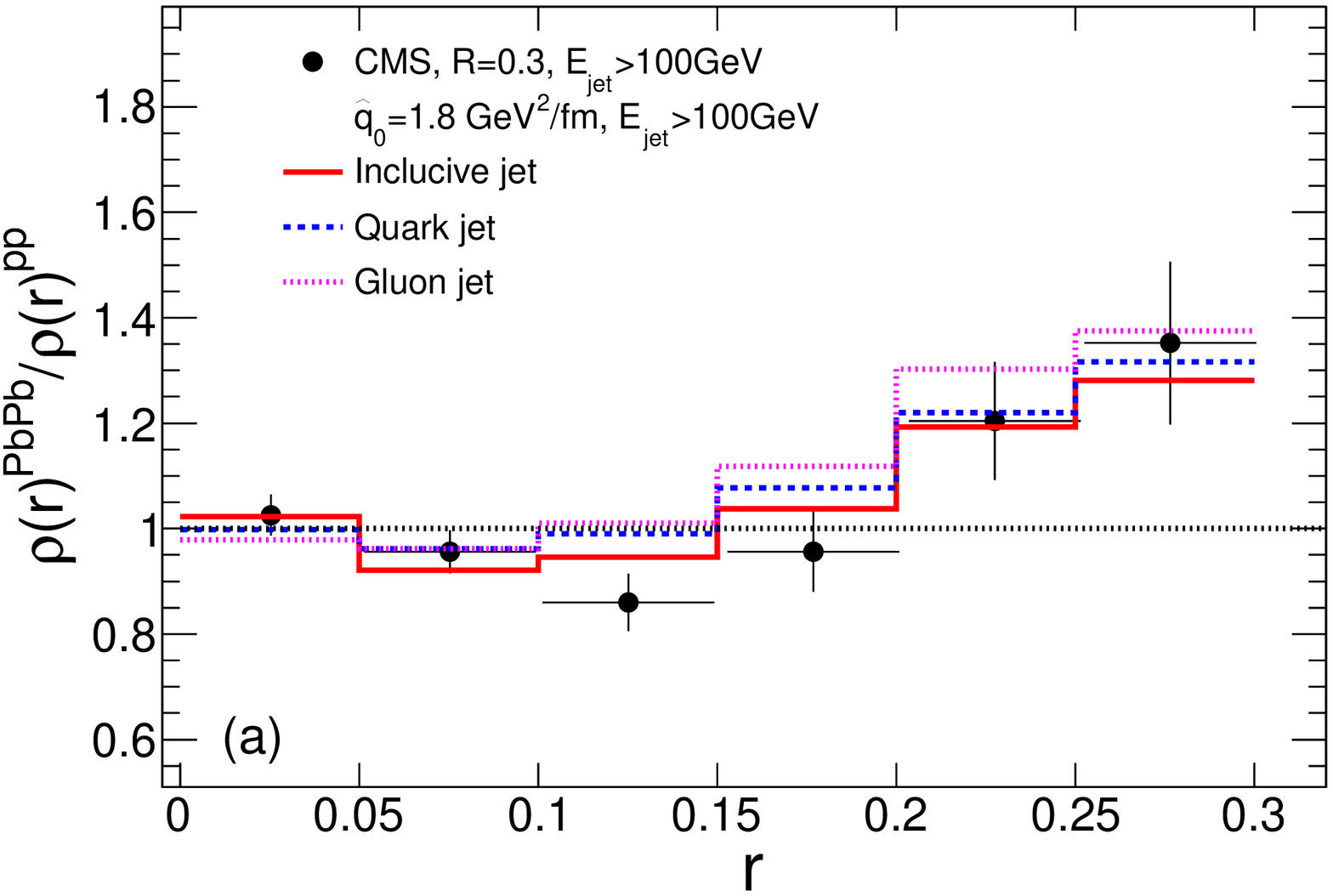}
     \includegraphics[width=1.0\linewidth]{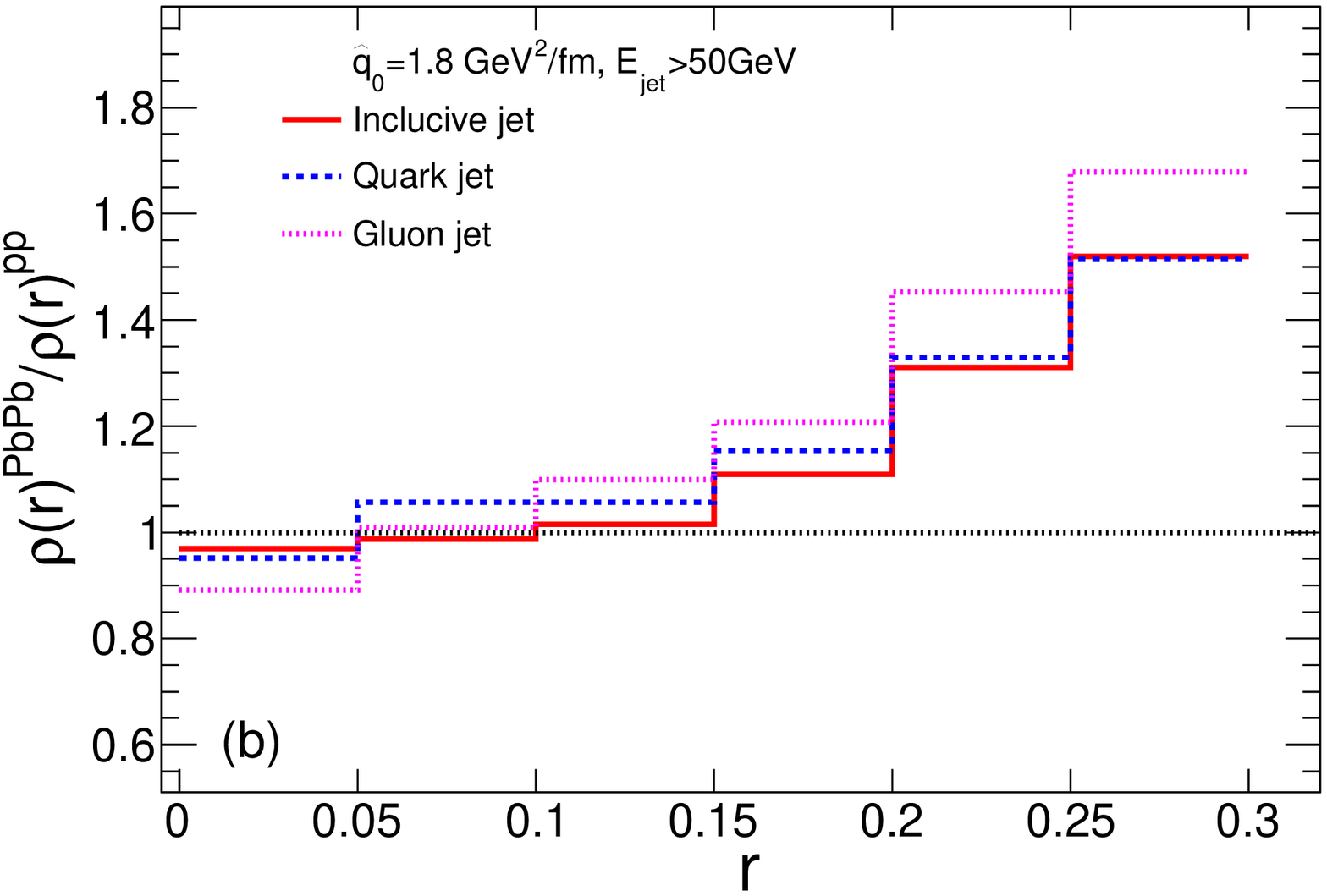}
  \caption{(Color online) The nuclear modification factor $R_{AA}^{\rho}(r) = \rho_{AA}(r)/\rho_{pp}(r)$ for the differential jet shape function of quark, gluon and inclusive jets in most central $0$-$10$\% Pb+Pb collisions at 2.76~ATeV: the upper panel (a) for $p_T>100$GeV and the lower panel (b) for $p_T>50$GeV.}
  \label{fig:rthoqg}
\end{figure}
\end{center}

As has been shown in Fig. \ref{fig:tho}, gluon jets are broader than quark jets in $p$+$p$ collisions. 
It is interesting to see whether the nuclear modifications of jet shape functions for gluon and quark jets are very different, given that gluon jets interact with the medium more strongly.
Such effect can be investigated experimentally by comparing the single inclusive jets and the jets tagged by isolated photons. 
In Ref.\cite{Chien:2015hda}, it was argued that the difference between quark and gluon jets can lead to a very different nuclear modification behavior for the photon-tagged jet shape function compared to Fig. \ref{fig:rtho}; it increases monotonously from small values of $r$ to large values $r$.
To check such effect, in Fig. \ref{fig:rthoqg} we show the nuclear modification factors of jet shape functions for quark and gluon jets. 
The results for two jet energy cuts are compared: $p_T>100$~GeV in the upper panel (a), and $p_T>50$~GeV in the lower panel (b).
We can see that the nuclear modification effect for the gluon jets is a little more than quark jets, but the difference is quite moderate. 
One may notice that the nuclear modification factor for inclusive jet shape function does not lie between those for quark and gluon jets in many $r$ bins. 
This is because the fractions of quark jets are larger in Pb+Pb than p+p collisions for jets with the same energies.
In this case, the nuclear modification factor $R_{AA}^{\rho}(r)$ of jet shape function for inclusive jets, as compared to those for quark and gluon jets, becomes sensitive to the ratio $\rho_g(r)/\rho_q(r)$ of quark's and gluon's jet shape functions, and the change of such ratio from p+p to Pb+Pb collisions; the latter is related to ratio of the nuclear modification factors of gluon's and quark's jet shape functions $R_{AA}^{\rho_g}(r)/R_{AA}^{\rho_q}(r)$.
Interestingly, we observe a monotonic behavior for the nuclear modification of jet shape function for both quark and gluon jets when a lower jet energy cut ($p_T > 50$~GeV) is used. 
In particular, a signifcant depeletion of the energy in very small values of $r$ is observed for lower energy jets ($p_T> 50$~GeV).
This very different from the case of higher energy jets ($p_T > 100$~GeV), for which there is much smaller modification of jet shape function at very small $r$. 
The above result tells that while jet-medium interaction does not affect much the inner core of very high energy jets, the medium can produce a sizable effect on the inner core of lower energy jets. 
We have also checked the jet energy dependence for the nuclear modification of photon-tagged jet shape function, and the result is quite similar to the case of single inclusive jets when the same jet energy cuts are used for the observed jets. 

\section{Roles of different jet-medium interaction mechanisms}
\label{sec:analysis}

In the previous section, we have presented the numerical results for different full jet observables, including the single inclusive jet $R_{AA}$, the dijet and photon-jet momentum imbalance, and the nuclear modification of jet shape function. 
The modification of full jets by the interaction with the medium is calculated from solving the coupled differential equations in Eq. (\ref{eq:dG/dt_q}) and Eq. (\ref{eq:dG/dt_g}) for the three-dimensional momentum distributon of the shower partons within the full jets. 
As has been mentioned, we have included several contributions during the full jet evolution, corresponding to various terms in the differential equations: the collisional energy loss and transverse momentum diffusion experienced by the shower partons, and the medium-induced radiative (splitting) processes. 
By turning on and off the corresponding terms in the evolution equations, we can investigate the contributions from different jet-medium interaction mechanisms on the propagation and modification of the full jets. 

\begin{center}
\begin{figure}[tbhps]
  \centering
     \includegraphics[width=1.0\linewidth]{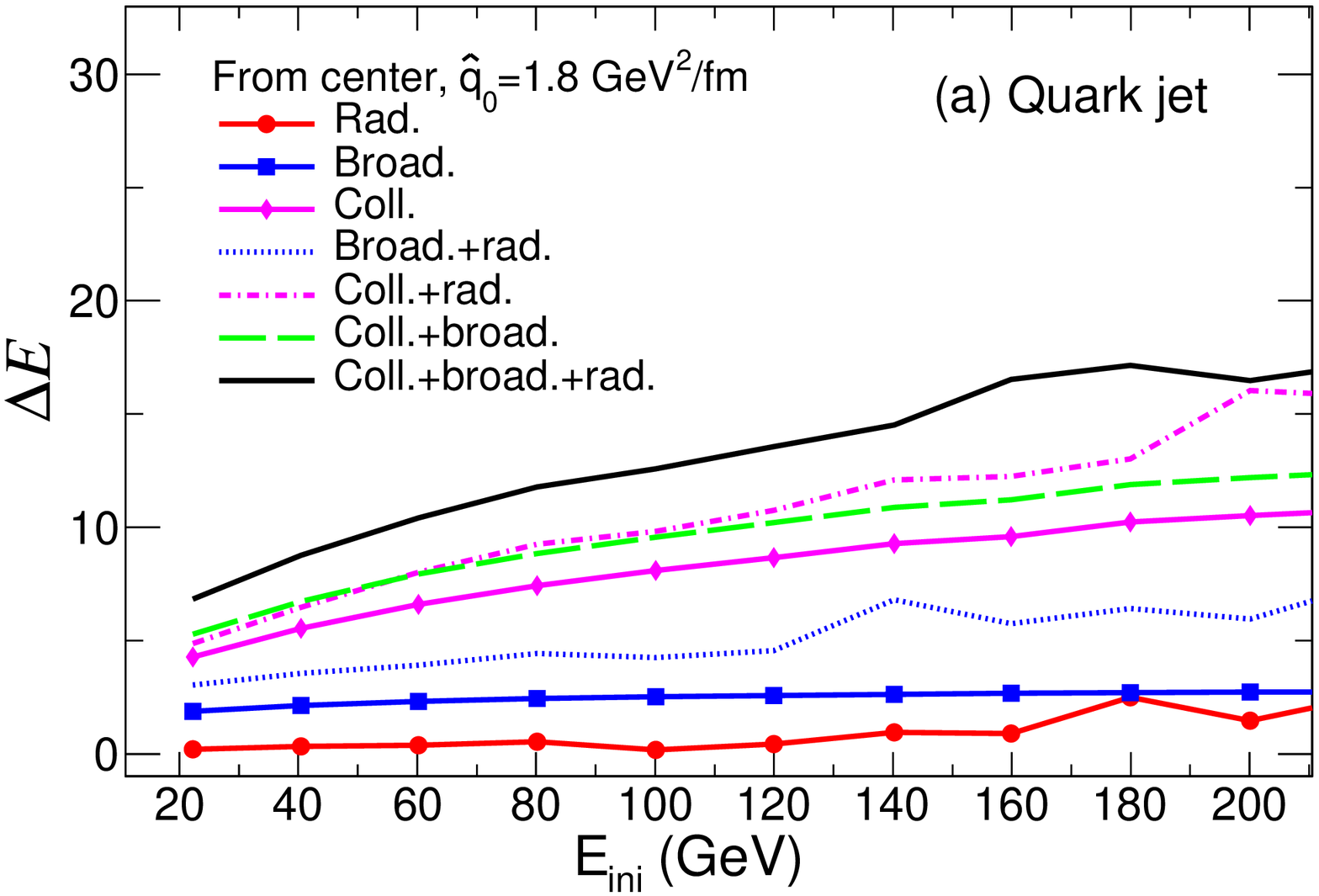}
     \includegraphics[width=1.0\linewidth]{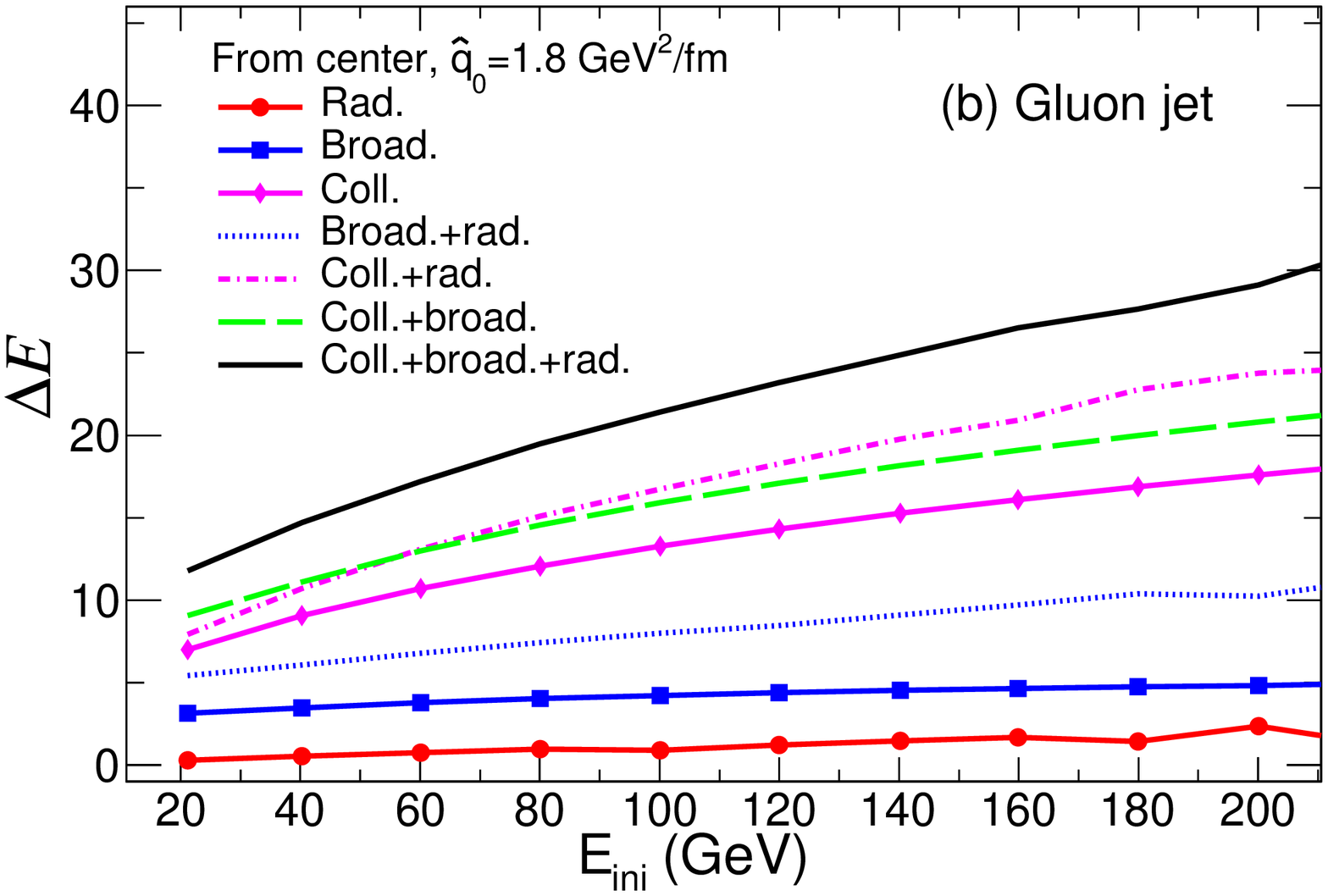}
  \caption{(Color online) The relative contributions from different jet-medium interaction mechanisms to the total full quark (a) and gluon (b) jet energy loss.
The jets are produced at the center of the medium created in most central $0$-$10$\% Pb+Pb collisions and propagte along $\phi=0$ direction.}
  \label{fig:eloss}
\end{figure}
\end{center}

In Fig.~\ref{fig:eloss}, we show the relative contributions to the total energy loss of the full jets from different jet-medium interaction mechanisms, namely the medium-induced radiation directly outside the jet cone, the transverse momentum broadening caused by the kicks from the medium constituents, and the collisional energy loss experienced by the shower partons. 
The upper panel (a) is for quark jets, and the lower panel (b) for gluon jets. 
In this figure, jets are initially produced at the center of most central $0$-$10$\% Pb+Pb collisions at 2.76~ATeV and propagate along $\phi=0$ direction in the transverse plane. 
Obviously, the gluon jets lose more energy than quark jets when they pass through the same medium. 
Also full jets with higher energies tend to lose more energy than lower energy jets when they pass through the same medium (note that the fractional energy loss descreases with the increase of jet energy).
Comparing different jet-medium interaction mechanisms, we observe that the most important contribution to the full jet energy loss is the collisional energy loss experienced by the shower partons. 
This can be understood from the fact that many soft partons at the periphery of the full jets are easily thermalized and absorbed by the hot and dense nuclear medium. 
Another observation is that the contribution from the medium-induced radiation directly outside the jet cone is very small, which is very different from the energy loss of leading parton. 
This is because most of the radiations from the full jet shower are collinear radiations which are inside the jet cone. 
The transverse momentum broadening via binary collisions with the medium constituents provide moderate contribution to total energy loss of the full jets. 
In the figure, we also show the combinational effects from the coupling of different interaction mechanisms. 
For example, although medium-induced radiation provides a very small direct contribution to the full jet energy loss, but it can greatly enhance the contribution from other mechanisms due to the increasing of the number of shower partons via the parton splitting process. 

\begin{center}
\begin{figure}[tbhps]
  \centering
     \includegraphics[width=1.0\linewidth]{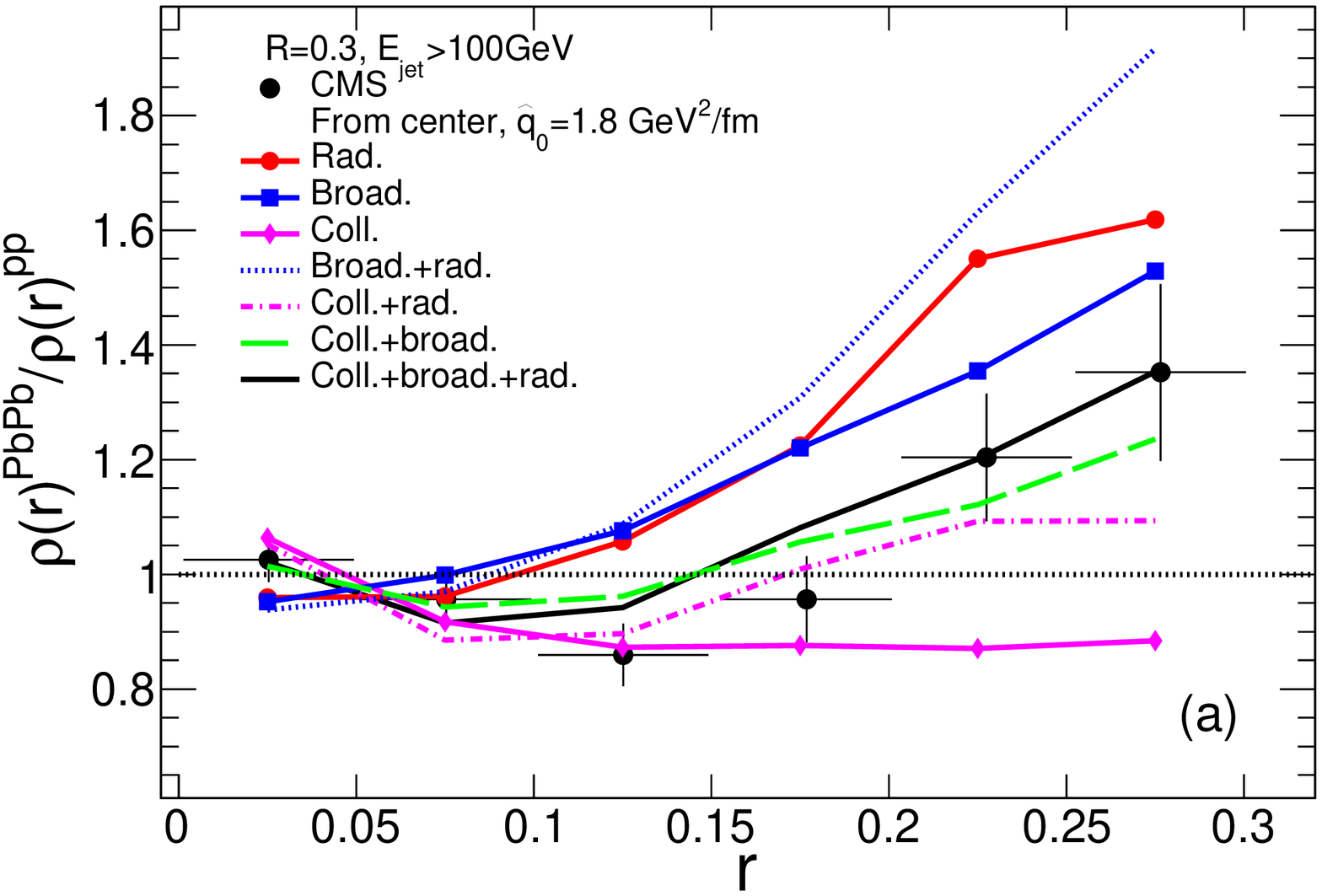}
     \includegraphics[width=1.0\linewidth]{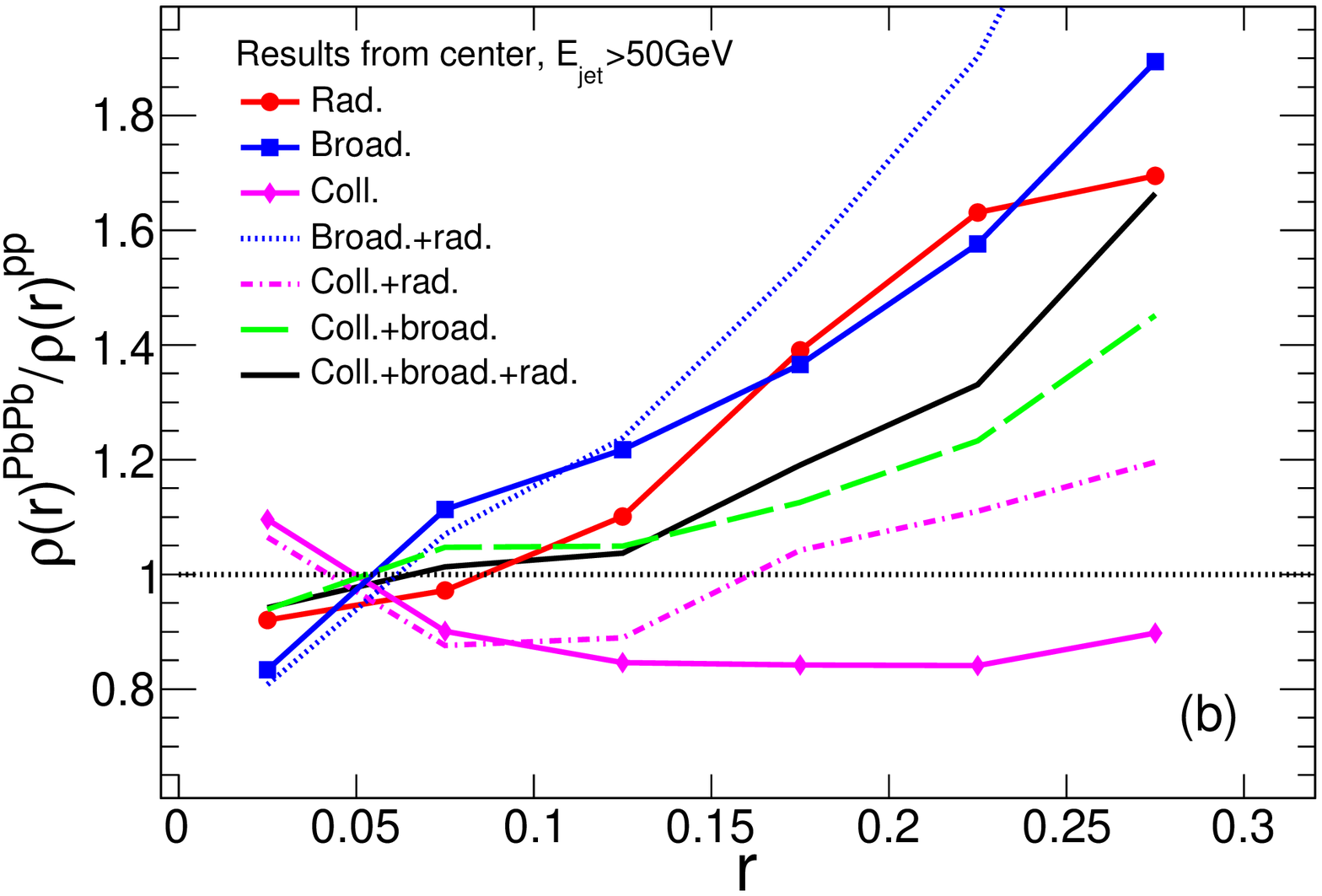}
  \caption{(Color online)
The relative contributions from different jet-medium interaction mechanisms to the nuclear modification of the differential jet shape function.
The jets are produced at the center of the medium created in most central $0$-$10$\% pb+pb collisions and propagte along $\phi=0$ direction. 
The upper panel (a) for inclusive jets with $p_T > 100$~GeV, and the lower panel (b) for $p_T > 50$~GeV.
}
  \label{fig:tho7}
\end{figure}
\end{center}

Now we analyze the effects from different jet-medium interaction mechanisms on the modification of jet shape function. 
As has been mentioned, the nuclear modification of jet shape function has a strong dependence on the jet energies. 
Therefore, in Fig. \ref{fig:tho7} we show the modication of jet shape function of single inclusive jet for two different jet energy cuts: the upper panel (a) for $p_T > 100$~GeV, and the lower panel (b) for $p_T > 50$~GeV.
We first look at the upper plot ($p_T > 100$~GeV).
One can see that while the medium-induced radiation provides very little direct contribution to full jet energy loss, it modifies jet shape function dramatically: it transports the energy of jet from the center to the periphery, and therefore leads to a suppression of jet shape function at the center and an enhancement at at the periphery. 
The transverse momentum broadening experienced by the shower partons has a similar effect. 
The collisional energy loss, however, has the inverse effect on the modification of jet shape function, i.e., it leads to a suppression at the periphery and an enhancement at the center. 
This means that the energy flowing to the medium from the leading parton is relatively smaller than that from the periphery of the full jets.
The reason is that the soft partons at the periphery of the full jets are easily to be absorbed by the medium, while the leading parton and the hard core of the full jets is much more difficult to be affected by the medium. 
The combination of the above three different effects leads to the same modification pattern as the results from the CMS Collaboration for $p_T>100$~GeV. 
Now we turn to the result for $p_T>50$~GeV (the lower plot). 
We can see that since the jet energy becomes smaller, the effects from all three different jet-medium mechanisms on the modification of jet shape function become a larger, both at the center and the periphery of the full jets.
One of the most interesting result is that the combination of three different mechanisms leads to a monotonic increase of the nuclear modification of the shape function for $p_T >50$~GeV, which is very different from the result for $p_T > 100$GeV (also see Fig. \ref{fig:rthoqg}).
This might be understood from the combination of two effects: the broadening effect of the jet shape increases and the change of the inner core of the full jets becomes easier, as one decreases the jet energies.

\begin{center}
\begin{figure}[tbhps]
  \centering
     \includegraphics[width=1.0\linewidth]{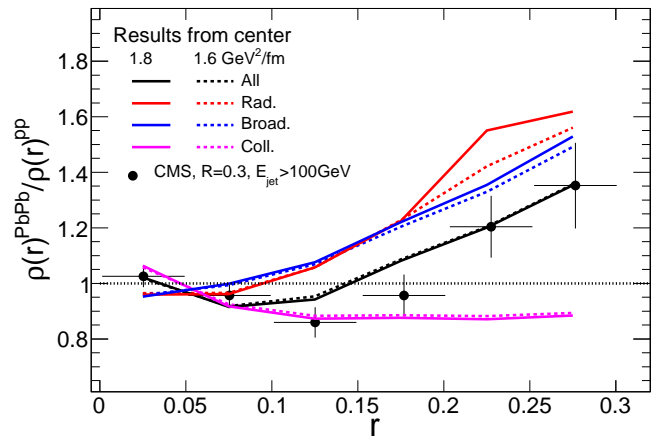}
  \caption{(Color online) The sensitivity of the nuclear modification of the differential jet shape function to the jet transport coefficient. 
The jets are produced at the center of the medium created in most central $0$-$10$\% pb+pb collisions and propagte along $\phi=0$ direction. 
The results for different jet-medium interaction mechanisms and their combination are compared. 
}
  \label{fig:tho72}
\end{figure}
\end{center}

Finally in Fig. \ref{fig:tho72}, we show the sensitivity for the nuclear modification of jet shape function to the value of jet transport parameter $\hat{q}_0$. 
We can see that with the increase of $\hat{q}_0$ value (as a result, the medium induced radiation, the transverse momentum broadening effect and the collisional energy loss all increases), the effects from three jet-medium interaction mechanisms on the nuclear modification of jet shape function increases when they are turned on individually. 
However, the final observed modification of jet shape function is a combinational effect from three different jet-medium interaction mechanisms, in particular, the modification from one interaction mechanism may be diminished or canceled out by another mechanism. 
Therefore, after turning on all three jet-medium interaction mechanisms, the nuclear modification effect of jet shape function actually becomes a little smaller when increasing the jet transport parameter $\hat{q}_0$. 
This means that the nuclear modification of jet shape function is very sensitive to the interplay of different jet-interaction mechanisms experienced by the full jet shower.

\section{Summary}
\label{sec:summary}

We have forumlated a set of coupled differential transport equations to study the evolution of full jet shower in quark-gluon plasma. 
The equations describe the evolution of the three-dimensional momentum distributions of quarks and gluons contained in the full jets, and incoporate three different jet-medium interaction mechanisms: medium-induced radiation, transverse momentum broadening and collisional energy loss, for both leading and radiated partons of the full jets.
During the evolution, we keep track of both the energies and the transverse momenta of all shower partons in the full jet, thus the medium modification of full jet energy as well as jet internal structure due to jet-medium interaction can be studied straightforwardly.
We apply our formalism to calculate the nuclear modification of single inclusive jet spectra, the momentum imbalance of photon-jet and dijet pairs, and the jet shape (at partonic level) in Pb+Pb collisions at 2.76~ATeV at the LHC by combining a realistic (2+1)-dimensional viscous hydrodynamic simulation to describe the space-time profiles of the hot and dense nuclear medium produced in the collisions.

By turning on and off different jet-medium interaction mechanisms, we have studied the effects from different mechanisms on the full jet energy loss and on the nuclear modification of the differential jet shape function. 
We find that the medium-induced radiation contributes very little to the direct full jet energy loss, but can increase the number of partons in the full jets. 
It can also transports the energy of full jet from the center to the periphery, which is similar to the transverse momentum broadening. 
Therefore, the medium-induced radiation and transverse momentum broadening both produce a suppression of jet shape function at the center and an enhanceat at the periphery. 
The collisional energy loss plays an important role in both full jet energy loss and the modification of the differential jet shower. 
Since the soft partons at the periphery of the full jets are easily thermalized by the medium and then the energy flows to the medium, thus it provides a significant contribution to total energy loss from the jet cone. 
It also leads to the suppression of jet shape function at the periphery and an enhancement at the center. 
Combing these three jet-medium interaction mechanisms, we obtain similar results as the CMS Collaboration for the nuclear modification of jet shape function.
We have also found that the nuclear modification of jet shape function is quite sensitive to full jet energies. 
For low energy jets, we observe a monotonically increasing behavior for the nuclear modification of jet shape function due to the increasing of the broadening effect as well as the easier modification of the inner part of the full jets, as compared to very high energy jets. 

In summary, various full jet observables in relativistic heavy-ion collisions can be explained by the interaction of full jets with the hot and dense medium created in the energetic collisions. 
The nuclear modification of jet shape function is very sensitive to the interplay of different jet-medium interaction mechanisms. 
Therefore it is important to include all relavant jet-medium interaction mechnisms for a comprehensive understanding of the medium modification of full jet production and jet shape function in relativistic heavy-ion collisions. 
In the future, we may improve our calculation in the following directions.
The inclusion of the energy dependence of various transport coefficients should be helpful to obtain a better description of the energy dependence on the full jet modification. 
The response of the medium to the energy and momentum deposition by the full jets may also affect the energy and internal structure of the reconstructed full jets \cite{He:2015pra}.
These are left for future effort.

{\bf Acknowledgments:} We thank X.-N. Wang for discussions. 
This work is supported in part by the Natural Science Foundation of China (NSFC) under Grant Nos.~11375072 and 11405066, Chinese Ministery of Science and Technology under Grant No. 2014DFG02050, the Major State Basic Research Development Program in China (No. 2014CB845404).


\bibliographystyle{h-physrev5}
\bibliography{GYQ_refs}

\end{document}